\documentclass{aa}
\usepackage{amsmath,a4wide,amssymb,calc,epsfig,longtable,graphicx}
\usepackage{natbib}\bibpunct{(}{)}{;}{a}{}{,}
\def\ldm{$\Lambda$CDM }

\newcommand{\mpch}{h^{-1}{\rm \mbox{Mpc}}}

\def\msolar{M_{\sun}}
\newcommand{\daverage}[1]{\left\langle #1 \right\rangle_{\rm u}}
\def\models{CDM: solid line; CHDM: short dashed line; $\Lambda$CDM: long dashed line.}
\def\d{{\rm d}}

\def\om{\Omega_{\rm m}}

\def\x{\vec{x}}

\def\V{\vec{V}}

\def\kpch{h^{-1}{\rm \mbox{kpc}}}

\def\p{\vec{p}}

\def\0{\vec{0}}
\def\ed{\ifmmode{{E}_d}\else{ ${E}_d$}\fi}
\def\fd{\ifmmode{{\mathcal F}_d}\else{ ${\mathcal F}_d$}\fi}

\def\api2{{$\Pi^{(2)}$}}
\def\api3{{$\Pi^{(3)}$}}
\def\api4{{$\Pi^{(4)}$}}
\def\L{{$\Lambda$CDM}}
\def\M{{CHDM}}
\def\gr{\kern 2pt\hbox{}^\circ{\kern -2pt K}} 

\newcommand{\omo}{\Omega_{{\rm m}}}
\renewcommand{\thefigure}{\arabic{figure}}
\renewcommand{\thetable}{\arabic{table}}

\def\ltsima{$\; \buildrel < \over \sim \;$}
\def\simlt{\lower.5ex\hbox{\ltsima}}
\def\gtsima{$\; \buildrel > \over \sim \;$}
\def\simgt{\lower.5ex\hbox{\gtsima}}
\begin{document}

\bigskip\medskip 

\title{The morphological and dynamical evolution\\ of simulated galaxy clusters}
\author{C. Beisbart\inst{1,2}
\and
R. Valdarnini\inst{3}
\and
T. Buchert\inst{4,5,1}}

\institute{Theoretische Physik, Ludwig-Maximilians-Universit\"at,
Theresienstr. 37, D-80333 M\"unchen, Germany
\and 
Astrophysics, Nuclear and Astrophysics Laboratory, Keble
Road, Oxford OX1 3RH, U.~K.
\and 
SISSA, Via Beirut 4, Trieste 34014, Italy
\and 
Theoretical Astrophysics Division, National Astronomical
Observatory, 2-21-1 Osawa Mitaka Tokyo 181-8588, Japan
\and 
D\'epartement de Physique Th\'eorique, Universit\'e de
Gen\`eve, 24 quai E. Ansermet, CH-1211 Gen\`eve, Switzerland}
\offprints{C. Beisbart, \email{C. Beisbart, beisbart@theorie.physik.uni-muenchen.de}}
\date{Received  0000,  Accepted 00000}  \abstract{  
We explore  the
morphological   and  dynamical   evolution  of   galaxy   clusters  in
simulations  using scalar and  vector-valued Minkowski  valuations and
the concept  of fundamental plane  relations.  In this  context, three
questions are of fundamental interest: 1. How does the average cluster
morphology  depend on the  cosmological background  model?  2.   Is it
possible to  discriminate between different  cosmological models using
cluster substructure  in a statistically  significant way? 3.   How is
the  dynamical state  of a  cluster,  especially its  distance from  a
virial equilibrium, correlated to  its visual substructure?  To answer
these  questions, we  quantify  cluster substructure  using  a set  of
morphological  order  parameters  constructed  on  the  basis  of  the
Minkowski  valuations  (MVs). The  dynamical  state  of  a cluster  is
described using  global cluster parameters: in certain  spaces of such
parameters fundamental band-like structures are forming indicating the
emergence of  a virial  equilibrium.  We  find that  the average
distances  from these  fundamental  structures are  correlated to  the
average amount of cluster  substructure for our cluster samples during
the  time  evolution.  Furthermore,  significant  differences show  up
between  the high-  and the  low-$\om$ models.  We  pay special
attention to the   redshift evolution of morphological characteristics
and find  large differences between  the cosmological models  even for
higher redshifts.
}
\titlerunning{Morphological evolution of galaxy clusters}
\authorrunning{C. Beisbart et al.}
\maketitle
\section{Introduction}
Galaxy clusters may be thought to constitute a sort of pocket guide to
our Universe:  although they are  small in comparison  to cosmological
scales,  they contain important  information about  the Universe  as a
whole. One line of thought  linking galaxy clusters and the background
cosmology  goes as  follows: according  to the  hierarchical scenario,
galaxy clusters were assembled through the merging of smaller objects,
which collapsed  first.  \citet{richstone:cluster} suggested  that the
cluster dynamical state  is related to its age,  which in turn depends
on average on the present  value of the cosmological density parameter
$\omo$.  If, finally,  the cluster dynamical state is  mirrored by its
substructure, one can establish  a link between cluster morphology and
the  background  cosmology {}\citep[``cosmology-morphology  connection
for  galaxy   clusters'',][]{evrard:morphology-cos}.   Therefore,  the
cluster substructure  may be a  powerful tool to study  the background
cosmology.     Summarising   the    results    of   the    theoretical
analyses~{}\citep[see also][]{bartelmann:clustercolla},  one can state
that in low-$\omo$  cosmologies the clusters should on  average show a
smaller amount of substructure than in high-$\omo$ models.  Since this
argument  oversimplifies  the complex  dynamical  situation in  galaxy
clusters,   it  has   to  be   complemented  using   simulations,  see
e.g.~{}\citep{evrard:morphology-cos}.  Note,  that we need  a thorough
definition   and  description   of  cluster   substructure   for  this
argument\footnote{From an observational point  of view, there is clear
evidence for  the existence of  substructure in galaxy  clusters, both
from                                                            optical
data~{}\citep{geller:cluster-substructure,dresslershectman,west:clusters,bird:mass}
and                             from                             X-ray
images~{}\citep{jones:imaging,boehringer:clusters,mohr:x-ray-substructure-methods}.}.\\
In this  context, it is  still a difficult  task to describe  both the
inner  cluster state and  the cluster  morphology quantitatively  in a
reliable  way.  --  In  this paper,  therefore,  we use  new tools  to
quantify cluster substructure as  well as the intrinsic cluster state.
We analyse  cluster simulations with these tools  and characterise the
substructure of  different cluster  components, its relation  to inner
cluster properties and the differences between cosmological background
models as traced by the averaged cluster substructure.  In particular,
we test the  theoretical assumptions behind the ``cosmology-morphology
connection''.
\\
So  far, various  methods have  been used  to quantify  the  amount of
substructure  in  galaxy  clusters.    In  the  optical  band  several
techniques~{}\citep{dresslershectman,west:clusters,bird:velocity}   use
the galaxy positions  and velocities.  Other methods are  based on the
hierarchical                                                 clustering
paradigm~{}\citep{serna:dynamical,gurzadyan:clusters},          wavelet
analysis~{}\citep{girardi:optical},  or moments  of  the X-ray  photon
distribution~{}\citep{dutta}.\\ X-ray  images of galaxy  clusters were
also used  to study substructure;  contrary to optical  clusters, they
are   scarcely   contaminated  by   fore-   and  background   effects.
\citet{mohr:x-ray-cluster} applied statistics based on the axial ratio
and        the        centroid        shift        of        isophotes
~{}\citep{mohr:x-ray-substructure-methods}   to  a   sample   of  {\it
Einstein  }   IPC  cluster  images.   \citet{buote:cluster-morphology}
introduced the power ratio method,  a technique based on the multipole
expansion  of the  two-dimensional potential  generating  the observed
surface               X-ray               brightness,              see
also~\citet{buote:cluster-morphologyII,buote:sensitivity,tsai:cluster-evolution,valdarnini:cluster}.
\citet{kolokotronis:searching}  studied  the  correlation
between substructures observed both  in the optical and X-ray bands.\\
Cosmological N-body simulations have  been used to test the dependence
of     cluster      substructure     on     different     cosmological
models~{}\citep{evrard:morphology-cos,mohr:x-ray-cluster,jing:clusters,thomas:clusters,west:rich}.
\citet{crone:cluster-substructure}   applied   different  substructure
statistics  to  galaxy  clusters  obtained in  different  cosmological
models   from   numerical  simulations.    They   conclude  that   the
``centre-of-mass shift'' is a  better indicator to distinguish between
different    models     than,    e.g.,    the     Dressler    Shectman
statistics~{}\citep{dresslershectman},   which    does   not   provide
significant                       results~{}\citep{knebe:substructure}.
\citet{pinkney:cluster_substructure} tested  several descriptors using
N-body simulations and recommended a battery of morphology parameters to
balance the disadvantages of different methods.
\\
So  far,   however,  a   unifying  framework  for   the  morphological
description of galaxy clusters is missing. Several aspects of cluster
substructure  have  to  be   distinguished  in  order  to  provide  an
exhaustive  characterisation.  Also  the  connection to  a  possible
cluster equilibrium has not yet been scrutinised.
\\
In    this     paper    we    apply     \emph{Minkowski    valuations}
(MVs)~{}\citep{mecke:robust,beisbart:querletter,beisbart:tensor_wup}
to  cluster  substructure  and  use \emph{fundamental  structures}  to
quantify  the  dynamical  state  of galaxy  clusters.   The  Minkowski
framework  provides  mathematically  solid and  unifying  morphometric
concepts, which can be applied to cluster data without any statistical
presumptions.    These   measures   distinguish  effectively   between
different aspects  of substructure and  discriminate between different
cosmological background  models.  Our interest  is both methodological
and  physical: on  the one  hand, we  are looking  for  an appropriate
method to  quantify cluster  substructure; on the  other hand,  we ask
physical  questions like: how  are the  Dark Matter  (DM) and  the gas
distribution related to each other?
\\
For   our  investigation,   we  employ   combined  N-body/hydrodynamic
simulations.  This  simulation technique is  particularly suitable for
our  purposes,  since it  traces  both the  dark  matter  and the  gas
component of a  cluster.  We construct relatively large  data bases of
cluster  images from  the simulations  which can  be compared  to real
cluster  images.\\ The  plan  of the  paper  is as  follows: after  an
explanation   of   the   simulations   and  cosmological   models   in
Sect.~\ref{sec:models},  we   give  an  introduction   into  Minkowski
valuations  in   Sect.~\ref{sec:mink}.   We  employ   these  tools  in
Sect.~\ref{sec:results} in  order to  compare the clusters  within the
different simulations.  An analysis  of fundamental plane relations is
presented   in  Sect.~\ref{sec:fp}.   We   draw  our   conclusions  in
Sect.~\ref{sec:con}.

\section{The cosmological models and the simulations}
\label{sec:models}
In order to investigate cluster substructure in different cosmological
models, a  data base of galaxy  clusters was generated on  the base of
TREESPH  simulations.    Three  background  cosmologies   were  chosen
differing both in  terms of the values of  the cosmological parameters
and  the power  spectra. We  restricted ourselves  to CDM  models; the
simulations      are     described      in     more      detail     in
\citet{valdarnini:cluster},  where also  a morphological  analysis was
done        using         the        power        ratios~{}\citep[PRs,
see][]{buote:cluster-morphology}.   We  extend  this work  in  several
directions,  e.g.   by  probing  the morphological  evolution  and  by
connecting cluster substructure and inner dynamical cluster state.
\\
Since observations indicate  that the curvature parameter $\Omega_{\rm
K}$   vanishes~{}\citep[see    for   instance][]{bernardis:flat}   we
considered  three  spatially  flat  cosmological  models,  namely  two
high-$\om$ models (a  standard Cold Dark Matter model --  CDM -- and a
model where the Dark Matter consists of a mixture of massive neutrinos
and Cold  Dark Matter -- \M) and  one low-$\om$ model (a  model with a
non-vanishing cosmological constant --  \L).  For the Hubble parameter
we chose  $h=0.5$ for the CDM and  the \M\ model, and  $h=0.7$ for the
\L\ model; here, as usual, the  Hubble constant is written in the form
$H_{\rm 0}= 100\, h\, {\rm \mbox{km}\, \mbox{s}}^{-1}\, {\rm \mbox{Mpc}}^{-1}$.
\\
With  respect to  the power  spectra comprising  the influence  of the
initial  matter  composition  on  structure  formation  we  adopted  a
primeval  spectral  index  of  $n=1$  and selected  a  baryon  density
parameter of  $\Omega_{\rm b} h^2 =  0.015$.  In the \M\  model we had
one massive  neutrino with  mass $m_\nu =  4.65\,$eV, yielding  a HDM
density parameter $\Omega_{\rm h} = 0.2$.  In the \L~ model the vacuum
contribution  to the  energy  density was  $\Omega_\Lambda  = 0.7$  in
accordance         with         recent         observations         of
Supernovae~{}\citep{perlmutter:supernovae}.   Therefore,  the  density
parameter of matter $\omo$ was $1$ for CDM and CHDM, and $0.3$ for \L.
Since we are dealing with galaxy clusters and need a fair number of them, all models were  normalised in order to match  the present-day cluster
abundance $n_{\rm c}(M>  M_{\rm c}) = 4 \cdot 10^{-6}  h^3$ for
$M_{\rm c} =
4.2                           h^{-1}                           10^{14}
M_\odot$~{}\citep{eke:cluster-evolution,girardi:optical}.  Using these
normalisations  only the \ldm  model is  consistent with  the measured
COBE quadrupole  moment at the $1  \sigma$ level.  In  order to reduce
the influence of cosmic variance  the same random numbers were used to
set the initial conditions for all cosmological models. Therefore
we roughly look at the same clusters in all cosmologies.
\\
The  cluster simulation technique  consisted of  two steps:  first for
each model a large collisionless N-body simulation was performed using
a P$^3$M code in a box of  length $L= 200\, a\, \mpch$, where $a$ is the
cosmological  scale factor being  one at  present day.   We considered
$N_{\rm p} =  10^6$ particles for the CDM and  \M~ models, each, while
$N_{\rm p}=84^3$  particles were  chosen for the  \L\ model,  the only
low-density  cosmology investigated here;  thus  the mass  of one
simulation   particle  is   approximately   equal in all
cosmological  models.  The  simulations  were run  starting  from  an
initial redshift $z_{\rm in}$, depending on the model {}\citep[for
more details see][]{valdarnini:cluster}, down to $z=0$.  At the final
redshift  we  identified  galaxy  clusters  using  a  friend-of-friend
algorithm in  order to detect  overdensities in excess of  $\simeq 200\,
\omo^{-0.6}$.  For our further analyses, we took into account only the
$40$   most   massive   clusters.  
\\
As     a      second     step     we      applied     a     multi-mass
technique~{}\citep{katz:hierarchical,navarro:cluster-simulation}:      for
each cluster  we carried  out a hydrodynamic  TREESPH simulation  in a
smaller  box  starting from  $z_{\rm in}$.   For  this  we identified  all
cluster particles within $r_{\rm 200}$ (where the cluster density is about
$200\,\omo^{-0.6}$  times the background  density) at  $z=0$.  These
particles were  backtracked to  $z_{\rm in}$ in the  original cosmological
simulation  box.  For  each  cluster  a cube  enclosing  all of  these
particles was constructed;  its size $L_{\rm c}$ was ranging  from $15$ to $
25 \mpch$.  A higher-resolution  lattice of $N_{\rm L}=22^3$ grid points then
was  set into  these  cubes.   Different lattices  were  used for  the
different mass components; to  avoid singularities these lattices were
shifted with  respect to  each other by  ${1/2}$ of the  grid constant
along  each  spatial  direction.   For  the \M\  simulations  the  hot
particles  bear   a  small  initial  peculiar   velocity  following  a
Fermi-Dirac  distribution  with  Fermi  velocity  $v_{\rm
0}=5(1+z_{\rm in})(10
{\rm eV}/m_{\nu})$ \mbox{km} s$^{-1}$.  For the gas particles we started with an
initial temperature  $T_{\rm i}=10^4{\rm K}$.  The TREESPH
simulation was then run using  all particles which lie inside a sphere
of radius $L_{\rm c}$ around the centres of the cubes.
\\
The gravitational softening parameters $\varepsilon$ were the same for
all clusters  within each simulation and cosmological  model.  For the
gas  particles they  were  chosen to  be $\varepsilon_{\rm  gas}=80,\,
100\,  ,60\,~{\rm \mbox{kpc}}$  for the  CDM, the~\M~,  and most  of  the \L~
clusters,  respectively.   However,  for  the five  most  massive  \L~
clusters  $\varepsilon_{\rm gas}$  was  set to  80$\,~{\rm \mbox{kpc}}$.   As
softening   parameters  for   the  Dark   Matter  particles   we  took
$\varepsilon_{\rm  d} =  200,231,125$ \mbox{kpc}  for  the CDM,  \M, and  \L~
model,  respectively.  For  the  simulation particles  we applied  the
scaling $\varepsilon_{\rm i} \propto m_{\rm i}^{1/3}$.  Note, that the
softening lengths  were fixed  within proper physical  space; however,
the redshift $z_{in}$ is chosen in  such a way, that the mean particle
separation is  always smaller than  the softening length.  The spatial
resolution  of   the  simulations  can  be  estimated   by  the  ratio
$\varepsilon_{\rm gas} / r_{\rm 200}$,  which never exceeds a value of
about  $0.04$.\\  The numerical  integrations  were  performed with  a
tolerance parameter  $\theta=0.7$ and using a  leap-frog scheme
for the time  integration; the minimum time step  allowed was $3\times
10^6$ years for the gas particles and $6\times 10^6$ years for the DM part.
Viscosity was  treated as in \citet{hernquist:tree} with  $\alpha = 1$
and  $\beta  =  2$.  The  effects  of  heating  and cooling  were  not
considered in the simulations. Tests assessing the quality of the
simulations are described  in {}\citet{valdarnini:cluster}.  We saved
numerical  outputs  at  different  redshifts, such  that  the  cluster
morphological  evolutions could be  investigated within  the different
models.
\\
Using  the  simulations  we   generated  cluster  images  which  mimic
observations in  a realistic  manner as follows:  the gas  density was
estimated on  a cubic  grid with a  grid constant of  $0.03~\mpch$ for
each model.  We took the square of this density at each grid point and
calculated  the approximate  integral of  $\rho^2$ along  the  line of
sight orthogonal  to a random plane  (it is the same  random plane for
all  clusters,  simulations,  and  redshifts),  with  $101\times  101$
pixels.  We  considered the cluster as  approximately isothermal, such
that   the   X-ray   emissivity   is   just   proportional   to   this
integral~{}\citep[see,    e.g.,][]{tsai:cluster-evolution}.    \\   We
applied the same method also to the DM particles; evidently, this does
not lead to a physically observable quantity.  However, in this way we
get the emissivity we
\begin{figure*}
\centering
\includegraphics[height=6cm]{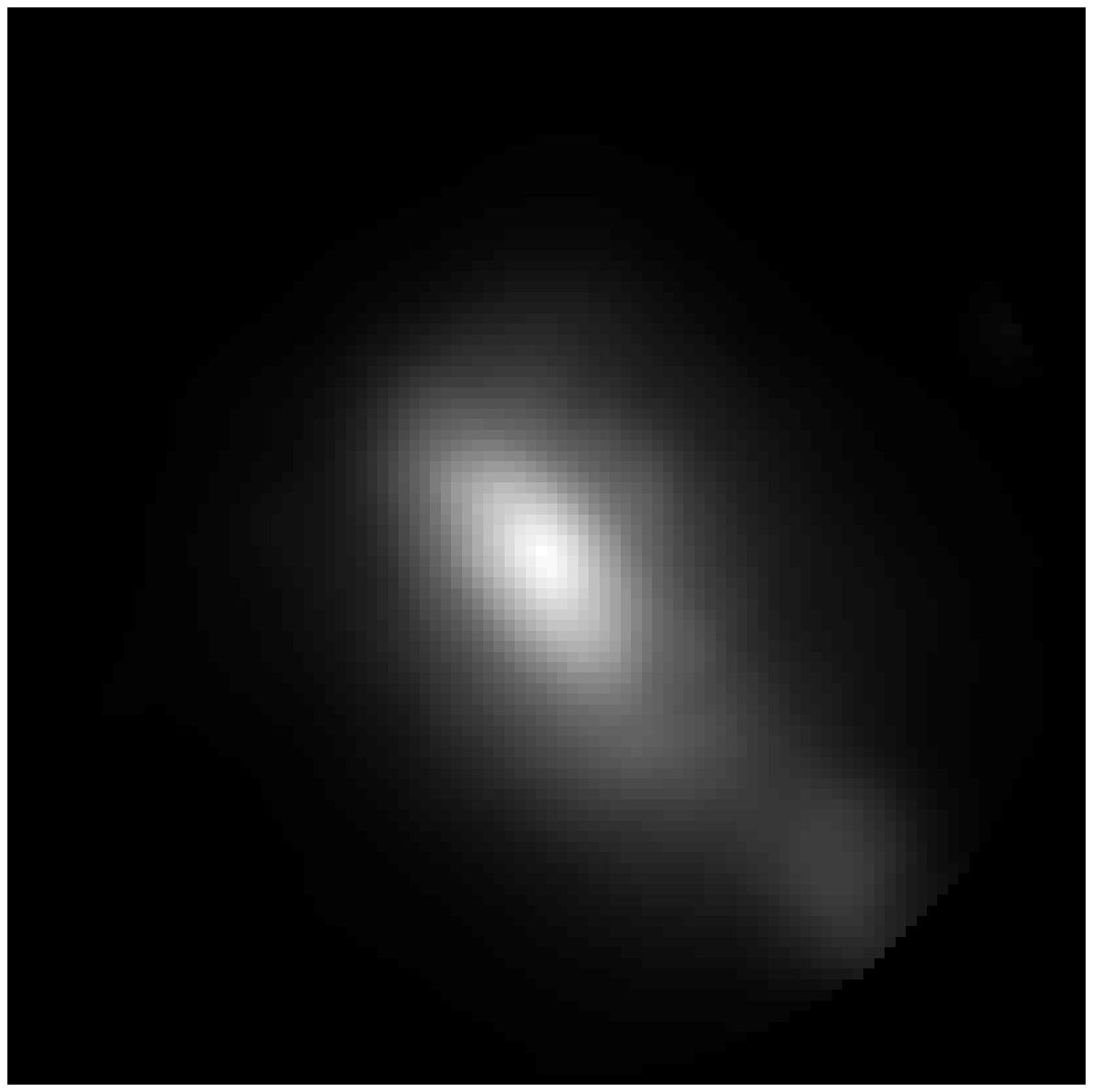}
\includegraphics[height=6cm]{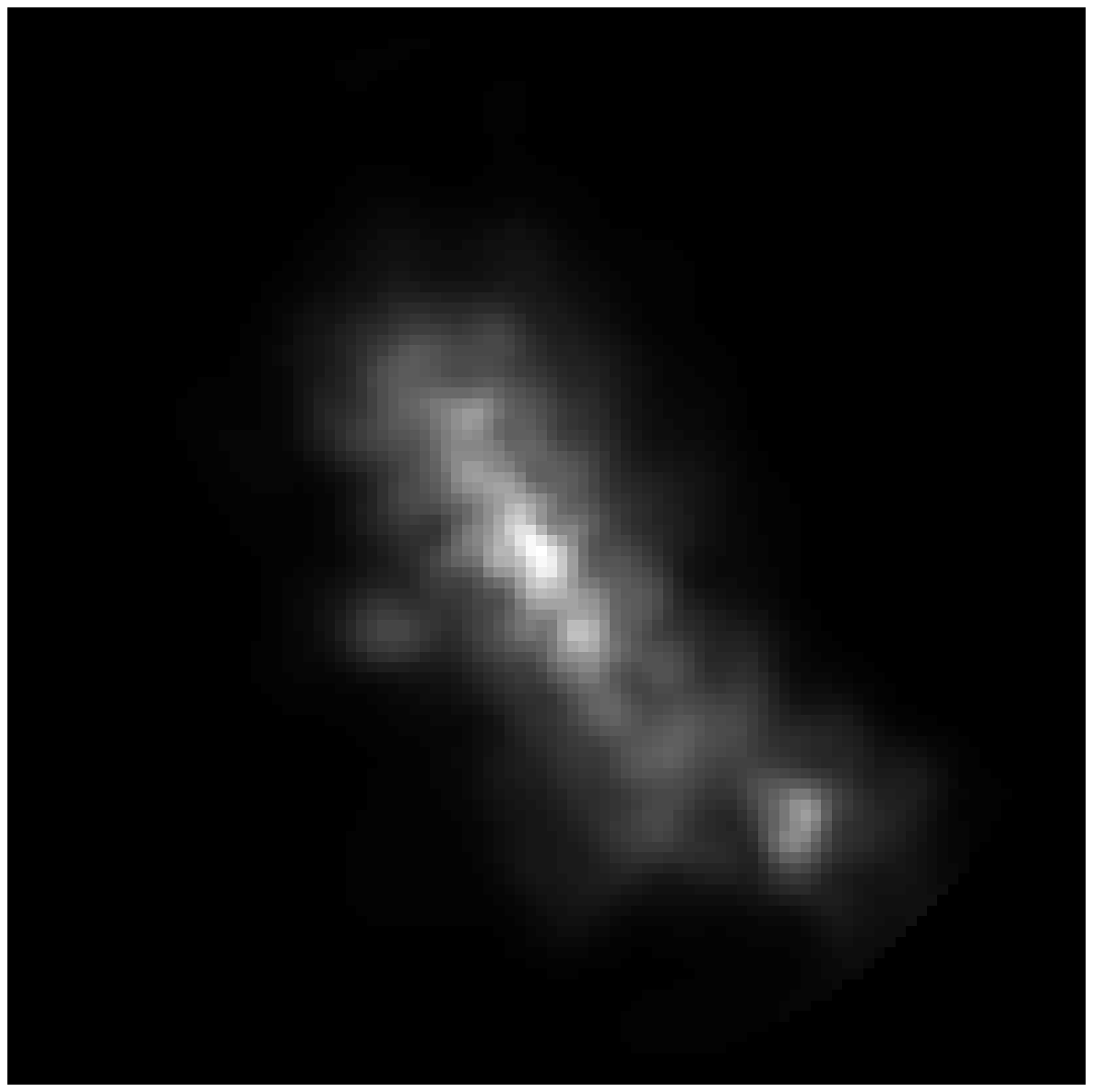}
\caption{Two  of the simulated  cluster images  to be  analysed.  They
show the  X-ray image of  one cluster in  the \ldm model  at redshift
zero (left  panel) and the corresponding  pseudo-DM-image, where the
gas density has  been replaced by the DM-density  (right panel). Both
images are  sampled within  a circle of  radius $1.5\mpch$  around the
peak of  the gas/DM-surface  brightness, respectively.  See  the main
text     for     a     further     description    of     the     image
construction.\label{fig:picture}}
\end{figure*}
would obtain  if the gas distribution  would trace the  DM (a constant
ratio between  gas and DM  distribution drops out in  our analysis).
We show  both an X-ray  and a DM-image  in Fig.~\ref{fig:picture}.
The  images are  analysed using  the Minkowski  valuations,  which are
described in the next section.
\section{Minkowski valuations}
\label{sec:mink}
The  \emph{Minkowski  valuations} (MVs)  provide  an elegant  and in  a
certain sense unique description of spatial data. They were introduced
into cosmology by~\citet{mecke:robust} and have been applied to answer
a  number of questions  regarding the  morphology of  the large-scale
structure,                                                         see,
e.g.~\citet{kerscher:abell,kerscher:fluctuations,schmalzing:cmb,sahni:shapefinders,schmalzing:webI,kerscher:pscz}.
So far, they were employed mainly in situations where perturbations of
a  homogeneous  background were  to  be  expected  and the  amount  of
clustering  had to be  quantified. For  galaxy clusters,  however, the
situation   is   different.    Galaxy   clusters   are   intrinsically
inhomogeneous systems, thus the main  issue is how far their structure
is away from  a symmetric and substructure-poor state  which does not
show the influence of recent mergers.
\\
For this reason, additionally  to the scalar Minkowski functionals, we
use vector-valued Minkow\-ski  valuations (also known as ``Querma{\ss}
vectors''), which  feature directional information\footnote{We reserve
the   name   ``Minkowski  functionals''   to   the  scalar   Minkowski
functionals. The  entirety of both scalar  and vector-valued Minkowski
measures is referred to as ``Minkowski valuations''.}. In this section
we give a short overview of both the scalar and the vector-valued MVs.
\\  For  a  general  approach,  let us  consider  patterns  $P,Q,...$,
i.e. compact sets within Euclidean space.  A morphometric (geometrical
and topological) description of  such spatial patterns is adequate, if
it  obeys  a  number  of  covariance  properties  specifying  how  the
descriptors  change  if the  pattern  is  transformed.  The  Minkowski
valuations are defined by three types of covariances.
\begin{enumerate}
\item  The first class  of covariance  refers to  motions in  space: a
morphological descriptor should behave  in a well-defined way, if the
pattern is  moved around in  space.  The scalar  Minkowski functionals
are motion-invariant,  whereas the vector-valued MVs transform
like vectors (\emph{motion equivariance}).
\item  Secondly,  the descriptors  should  obey  a  simple rule 
specifying how they sum up  for combined patterns whose set union
is  constructed.  Both  classes of  Minkowski valuations  $V$  obey the
\emph{additivity}:
\begin{equation}
V(P\cup Q) =V(P) + V(Q)- V(P\cap Q)\;\;\;.
\end{equation}
\item Moreover, \emph{continuity} states that a descriptor
should   change    continuously   if   the    pattern   is   distorted
slightly\footnote{
  A closer  analysis shows that  this requirement can be  only imposed
  for convex bodies.  }.
\end{enumerate}
These simple properties  already define the MVs, since  -- at least on
the convex ring\footnote{
  The convex ring comprises all finite unions of convex bodies.  } the
Theorem     of     Hadwiger~{}\citep{hadwiger:vorlesung,hadwiger:vekt}
guarantees that  in $d$ dimensions,  only $(d+1)$ of such  measures --
either scalar-  or vector-valued  ones -- are  linearly independent.
\\ Because of the structure of our cluster data, we concentrate on the
case  of  $d=2$.   Here,  the  Minkowski  functionals  have  intuitive
meanings:  $V_0$ is  the surface  content  of the  pattern, $V_1$  its
length of  circumference, and  $V_2$ its Euler  characteristic $\chi$,
which counts the components of the cluster and subtracts the number of
holes.  Note,  that all  of these functionals  can be expressed  as an
integral: $V_0$  obviously is  the two-dimensional volume  integral of
the  pattern, $V_1$  is  its (one-dimensional)  surface integral,  and
$V_2$  weights each  surface element $\d S$ with  the local  curvature
$\kappa$.   This  integral
representation   is  also  valid   for  the   vector-valued  Minkowski
valuations; in this case, additionally, the integrals are weighted with the position
vector; therefore, the Querma{\ss}  vectors are spatial moments of the
scalar   Minkowski  functionals\footnote{More   precisely,   they  are
first-order  moments.  Higher  moments are  investigated  as Minkowski
tensors in forthcoming work~\citep{beisbart:tensor_wup}.}. Summarising
we consider the following measures:
\begin{alignat}{3}
 V_0   &=  \int_P \d V\;\;\;,      \quad              &\V_0 &=  \int_P \d V \x   \;\;\;,            \\ \notag 
 V_1   &=  \int_{\partial P} \d S\;\;\;,\quad    &\V_1 &= \int_{\partial P}\d S\x\;\;\;,\notag \\ 
 V_2   &=  \int_{\partial P} \d S \kappa\;\;\;, \quad  & \V_2 &=
 \int_{\partial P}\d S \kappa \x \;\;\;, \notag    
\end{alignat}
where $\kappa$ refers  to the local curvature. --  It is convenient to
divide  the vectors  by the  corresponding  scalars to  arrive at  the
(curvature) centroids\footnote{
  In  the   following  we   speak  of  ``centroids''   or  ``curvature
  centroids'',  the  latter  term  becomes more  plausible  in  higher
  dimensions, where almost all centroids are connected to curvature.
  }:
\begin{equation}
\p_i \equiv \frac{\V_i}{V_i}\;\;\;.
\end{equation}
The   centroids  localise   individual   morphological  features.   In
particular,  they coincide  with the  symmetry centre  for spherically
symmetric  patterns.   On the  other  hand,  centroids constituting  a
finite triangle indicate the presence  of asymmetry. \\ Note, that the
MVs  obey covariance  properties  with  respect to  a  scaling of  the
pattern $P$, too: If we scale the pattern by $\alpha>0$ to get $\alpha P
= \{ \alpha \x|\x\in  P\}$, the scalar Minkowski functionals transform
like  $V_i (\alpha P)  = \alpha^{d-i}V_i  (P)$; the  vectors transform
like $\V_i  (\alpha P) =  \alpha^{d-i+1}\V_i (P)$.  This  is important
since  we  sometimes have  to  compare  data  of different  size.   \\
Obviously, cluster  images are  not patterns in  the above  sense, but
rather consist  of galaxy positions  or pixelised maps  reflecting the
surface brightness  within a  certain energy band.   Thus, we  have to
construct  patterns from  the  cluster  data.  Here  we  use the  \emph{
excursion  set approach}  where we  smooth the  original data  using a
Gaussian  kernel  in order  to  construct  a  realization of  a  field
$u(\x)$.  The excursion sets
\begin{equation}
M_w  = \{ \x| u(\x) > w\}\;\;\;.
\end{equation}
are analysed  with the aid  of the Minkowski valuations.   Varying the
density  threshold $w$  allows us  to probe  different regions  of the
cluster.  The  smoothing both reduces  possible noise and picks  out a
scale  of  interest or  resolution  beneath  which substructure  is no longer
resolved.\\  The MVs, represented as functions  of the density
threshold, contain very detailed information.  In this paper, however,
we  want  to compare  cluster  images drawn  from  a  larger base  of
simulated galaxy clusters.  We are therefore interested in the average
morphological cluster evolution in different cosmologies.  In order to
condense  the detailed information  present in  the MVs,  we construct
robust  structure functions  which  allow us  to  compare clusters  of
different  size statistically.  \\  This can  be done  by integrating
over  the  density  thresholds  and  weighting with  functions  of  the
Minkowski  valuations. We  define  an average  over different  density
thresholds via
\begin{equation}
\daverage{f} \equiv  \frac{1}{u_{\rm max}-u_{\rm min}} \int_{u_{\rm
min}}^{u_{\rm max}}  \d u f \;\;\;.\label{eq:daverage}
\end{equation}
Here, we  consider three classes of robust  structure functions, which
feature different aspects of the substructure and span a morphological
phase  space.    We  distinguish   the  following  three   classes  of
substructure, which  are quantified using  one or more \emph{structure functions}:
\begin{enumerate}
\item the  {\it clumpiness} $C\equiv\sqrt{\daverage{(\chi-1)^2}}$ is a  measure of the
number of subsystems in a cluster;
\item     the     \emph{shape     and     asymmetry     parameters}
($A_0\equiv\daverage{{\rm Area}(\triangle  (\p_j))} $, $A_1=\daverage{
{\rm perimeter}(\triangle (\p_j))}$) refer  to the degree of asymmetry
and the  global shape  of the cluster  (``is the cluster  spherical or
elongated?''); here we use the  fact that curvature centroids which do
not coincide within  one point indicate the presence  of asymmetry. In
particular, the size of the triangle $\triangle (\p_j)$ which connects
the curvature centroids (measured by its area and perimeter) serves as
a measure of the asymmetry present within the cluster.
\item  the  {\it  shift  parameters} $S_{i=0,1,2}\equiv\sqrt{\daverage{|\p_i-\daverage{\p_i}|^2}}$  account  for  the
variation or  shift of morphological properties in  a quantitative way
by  considering  different density  thresholds  (these parameters  are
generalisations  of  the frequently  used  centre-of-mass shift  and
centroid                                                     variation,
see~\citet{crone:cluster-substructure,mohr:x-ray-substructure-methods,mohr:x-ray-cluster})\footnote{It
turns out that $S_2$ is closely related to the clumpiness. -- For the
sequel we concentrate on the structure functions $C$, A$_1$, and
S$_1$. We successfully tested  the robustness of these latter structure functions.}.
\end{enumerate}
 An effective method to  calculate Minkowski valuations from pixelised
maps can  be developed in analogy  to~\citet{schmalzing:beyond} and is
given in ~\citet{beisbart:quer}.
\section{Cluster substructure and the background cosmology}
\label{sec:results}
\begin{figure*}
\centering
\includegraphics[height=4.2cm]{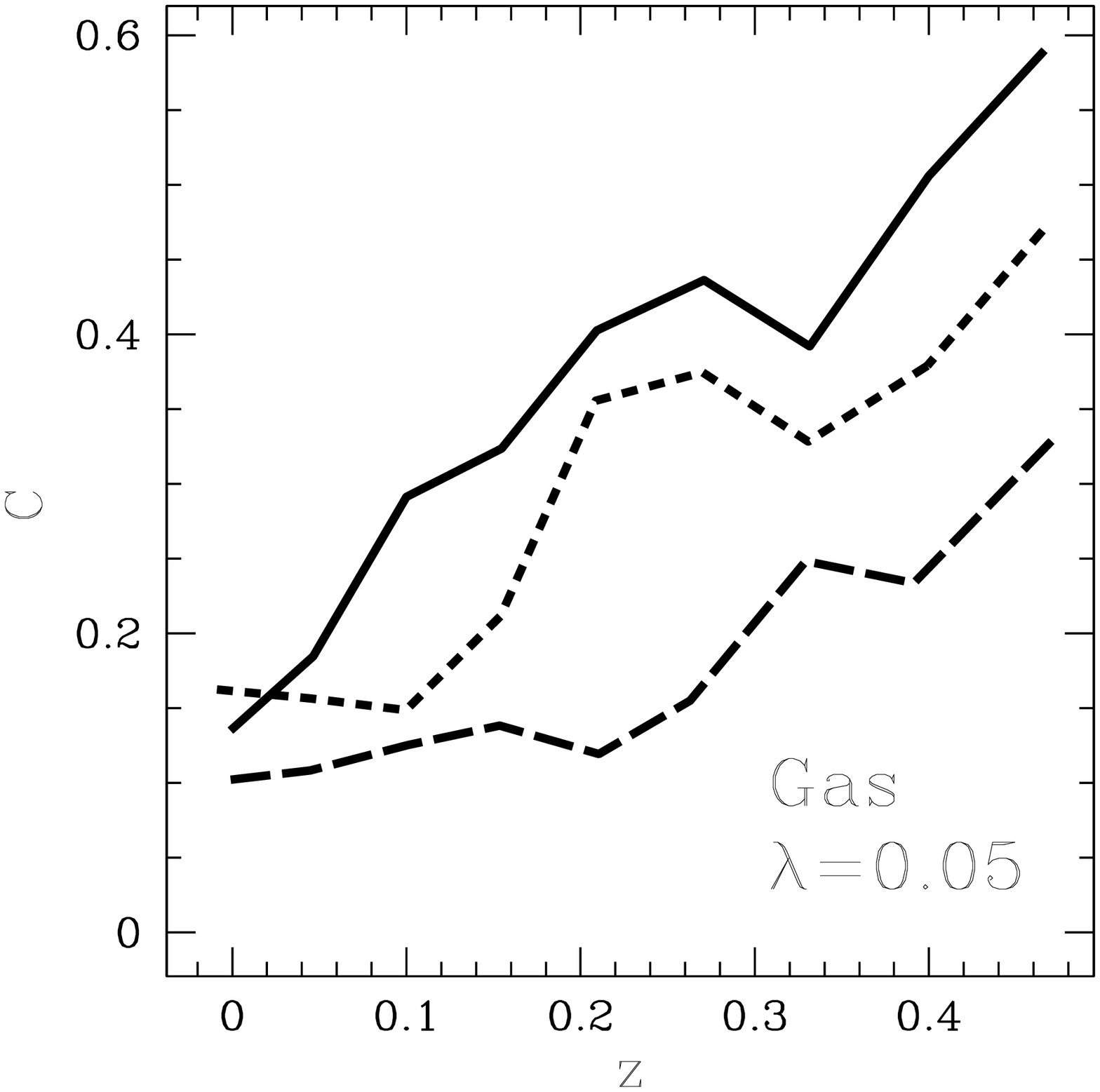}
\includegraphics[height=4.2cm]{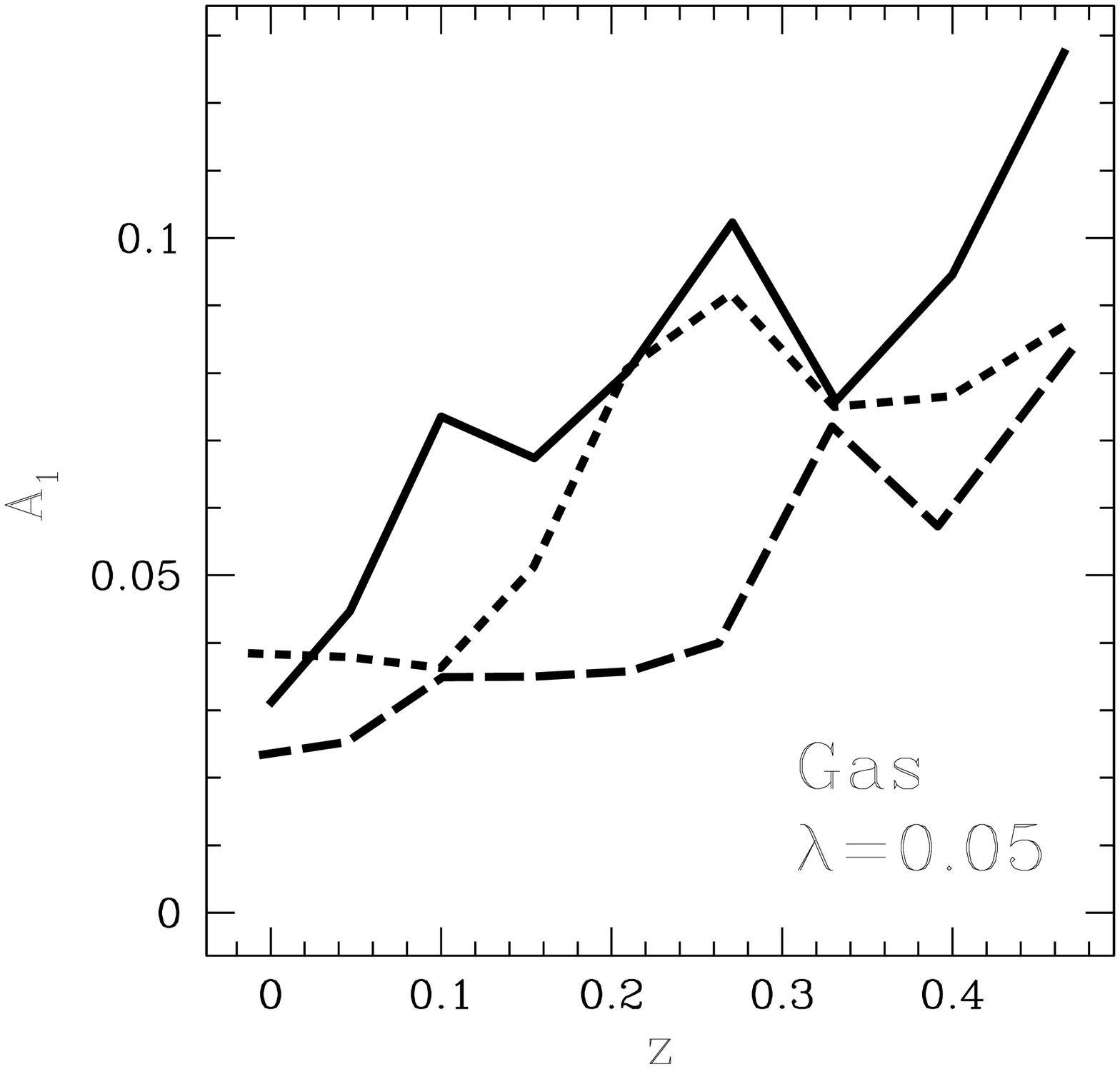}
\includegraphics[height=4.2cm]{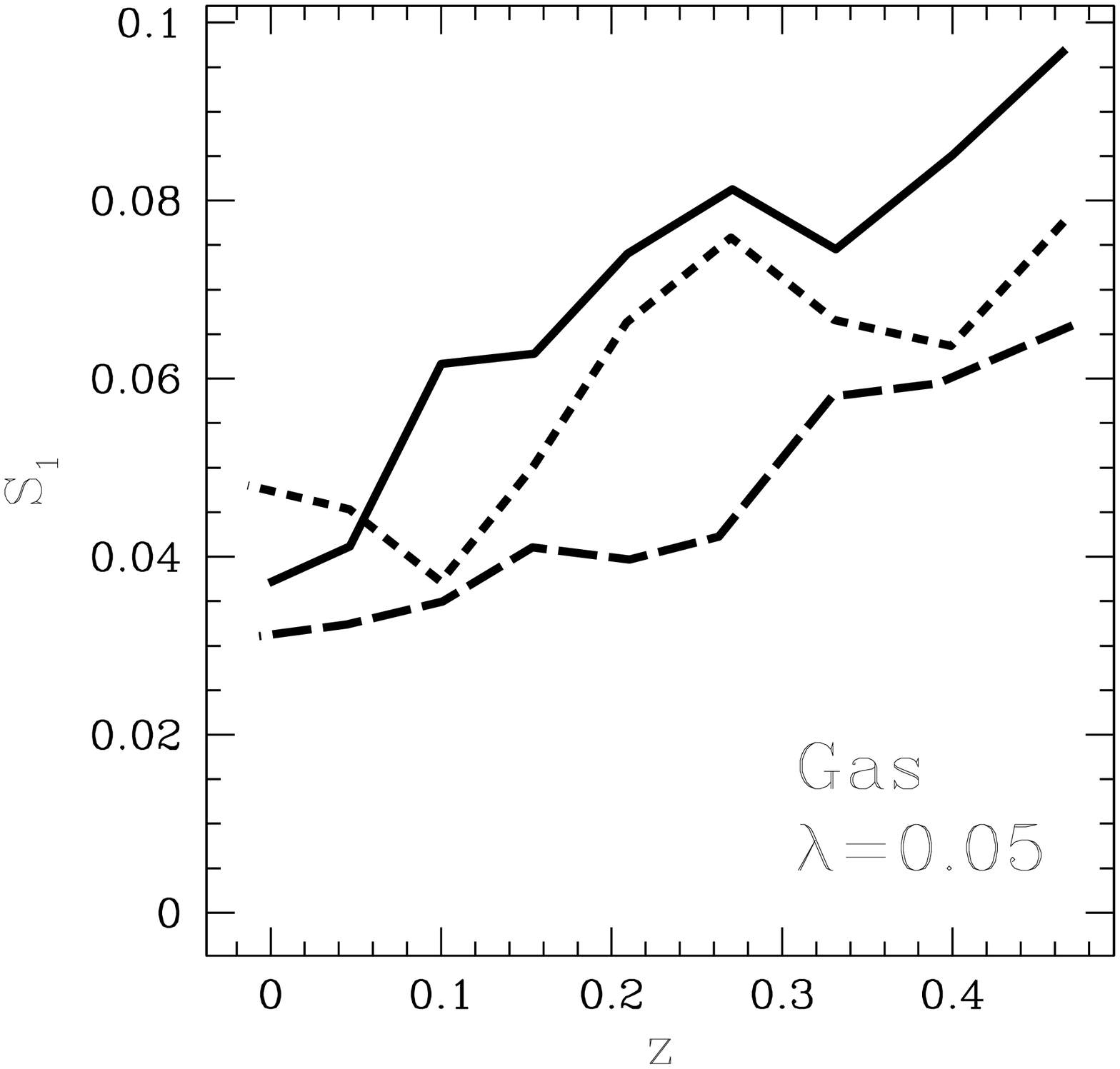}
\includegraphics[height=4.2cm]{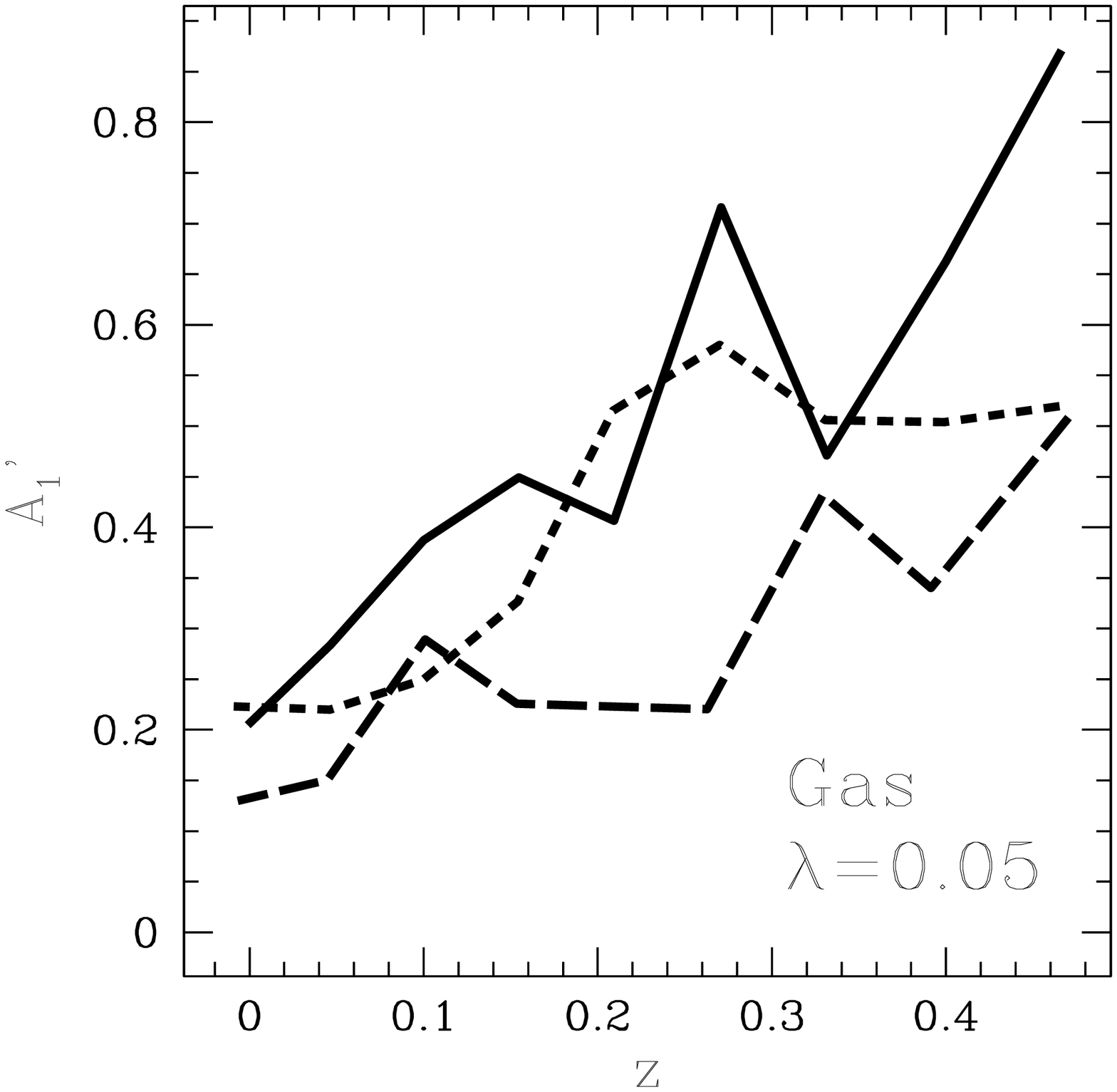}
\includegraphics[height=4.2cm]{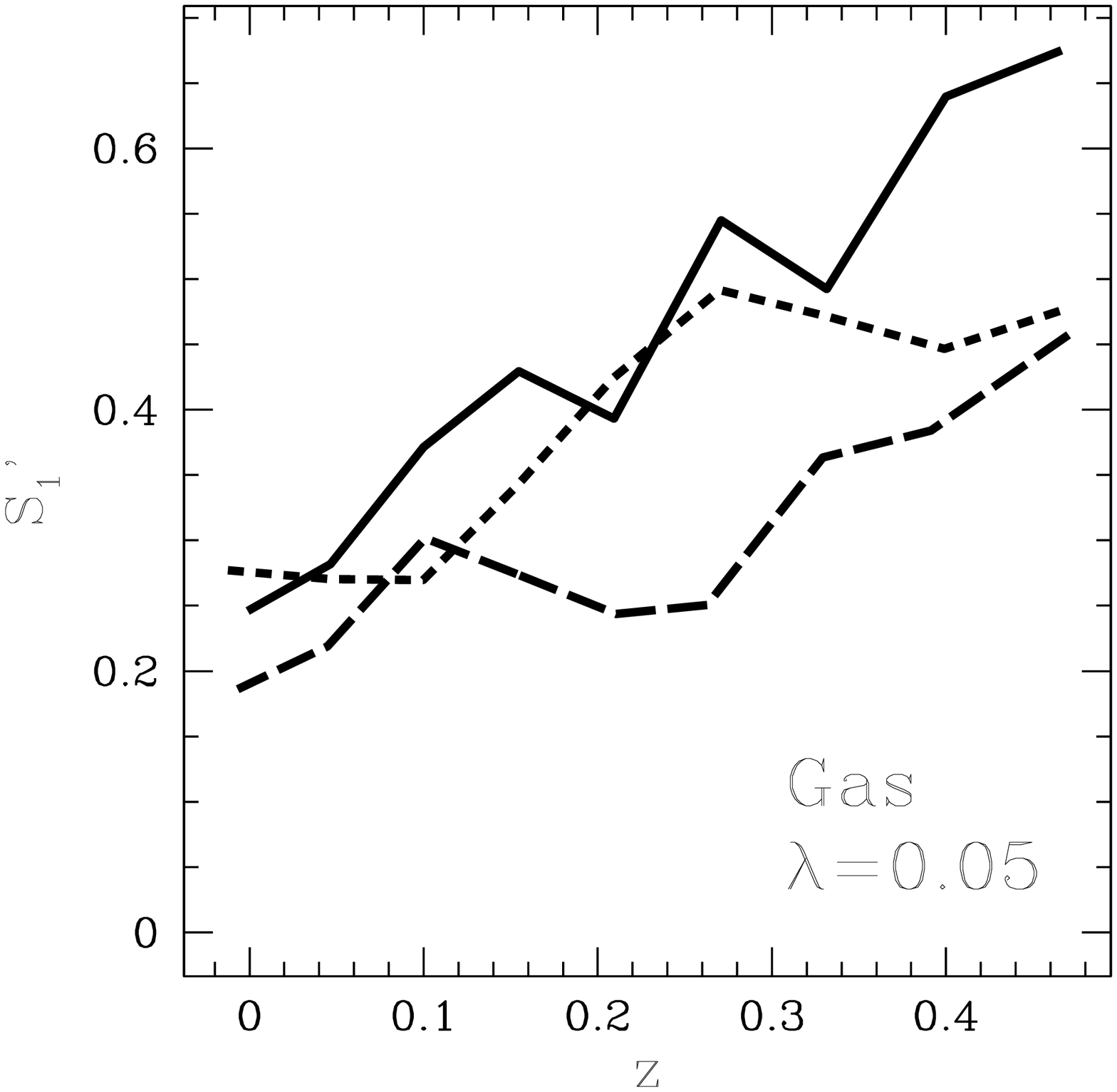}
\caption{ The averaged morphological evolution of the galaxy clusters
in our three cosmological models. The ensemble-averaged structure functions \emph{clumpiness},
\emph{shape and asymmetry},  and \emph{shift of morphological properties}
were determined from the  X-ray luminosity for  a smoothing
length  of  $\lambda=0.05\mpch$ and are shown vs.  redshift (first row).  \models\;   We consider  a  spherical
window  with radius  $r_{\rm w}=1.4 \mpch$;  C is  dimensionless;  A$_1$ and
S$_1$ are given in units  of $(\mpch)^2$ and $\mpch$, respectively. In
the second row we show  the A$_1^\prime$ and S$_1^\prime$, where A$_1$
and S$_1$ were  scaled to cluster size (an estimate  of the half light
radius) in order to get dimensionless parameters.} 
\label{fig:compg0.05}
\end{figure*}

To  start  our  analysis  of  the simulated  clusters,  we  probe  the
connection   between  the   background  cosmology   and   the  cluster
morphology.   Since in  this  case  not so  much  the substructure  of
individual clusters rather than the mean morphology is of interest, we
define  \emph{cluster  samples}  consisting  of all  clusters  at  one
redshift  within one  model (unless  otherwise stated  we  analyse one
random  projection per  cluster)\footnote{On  account of  applications
below, we  discarded a few  clusters which showed  obvious pathologies
such as a strong bimodality. We  have $39$ clusters for the CDM model,
$37$ for the CHDM and $35$ for the \ldm model.}.  In order to trace the mean
substructure evolution,  we average  the structure functions  over all
clusters within one sample.
\\
What is  the amount  of cluster substructure,  and how does  it evolve
within the  three cosmological models?  -- The  simulated clusters are
``observed'' within  a quadratic window  of $3\mpch$ width  centred at
the peak  position of the  surface brightness.  The data  are smoothed
using a Gaussian  kernel with different smoothing lengths  in order to
reduce the sensitivity  to noise and to probe  different scales of the
substructure.  We concentrate on  intermediate values of the smoothing
scale $\lambda$  ($\lambda\sim0.05-0.15\mpch$, \citet{cen:pro} employs
values of the same order of magnitude)\footnote{ A smoothing length of
$0.05\mpch$  is  smaller or  equal  than  the gravitational  softening
length for  the gas and not much  smaller than that one  for the DM.}.
To define  the cluster on the  image, we draw circles  around the peak
with radii $r_{\rm w}=0.8,1.0,1.2,1.4\mpch$ (this definition is in the
spirit  of   Abell's  cluster  identification  in   the  optical,  see
{}\citealt{abell:clusterkatalog},  we  call  this circle  the  cluster
window) and neglect the rest.   The integration limit $u_{\rm min}$ in
Eq.~\eqref{eq:daverage} is  chosen to  be twelve times  the background
which  is  determined  from  the   rest  of  the  image  similarly  as
in~\citet{boehringer:northern},         $u_{\rm        max}$        in
Eq.~\eqref{eq:daverage} is the maximum cluster surface brightness.
\\
In Fig.~\ref{fig:compg0.05} we show results for  the X-ray cluster
morphology within  the three models having applied  a smoothing length
of  $0.05~\mpch$.  A  couple of  things are  obvious at  first glance:
there is a significant  difference between the high-$\om$ models (CDM
and CHDM) and the  low-$\om$ model ($\Lambda$CDM).  These differences
are visible in  most of the structure functions  and are in accordance
with the  theoretical expectations: the low-$\om$ model  shows by far
less  substructure than  the other  two models  -- at  least  for most
redshifts investigated here.  The CDM and CHDM models, however, do not
seem  to be distinguished  well.  Therefore,  the morphology-cosmology
connection  is mainly  sensitive  to the  values  of the  cosmological
parameters,  but performs poorly  in discriminating  between different
power  spectra.    The  clumpiness  is   particularly  sensitive.   --
Regarding  the redshift evolution,  a clear  trend is  visible towards
more relaxed and substructure-poor clusters. In particular, there are
also  large morphological  differences  between the  models at  higher
redshifts.   \emph{Morphological  evolution  of galaxy  clusters}  may
therefore serve as a more sensitive test than the present-day cluster
morphology.  Note, that the  averaged morphology evolution still looks
relatively  spiky.  The  reason is  that for  individual  clusters the
evolution  of  the structure  functions  proceeds  in a  discontinuous
manner, when subclumps enter the cluster window. Therefore, one has to
average over several clusters in  order to get a typical morphological
cluster evolution.
\\
The structure functions A$_i$ and S$_i$ have a dimension and therefore
 quantify  the absolute  amount of  substructure.  To  investigate the
 substructure relative  to cluster size, we normalise  A$_1$ and S$_1$  to the
 individual cluster
 size  estimated via  the two-dimensional half-light  radius around  the peak  of the
 X-ray surface   brightness.    As   visible   from  the   bottom   row   of
 Fig.~\ref{fig:compg0.05},   the  qualitative   evolution   and  the
 differences between the models are similar as before.
\begin{figure*}
\centering
\includegraphics[height=4.9cm]{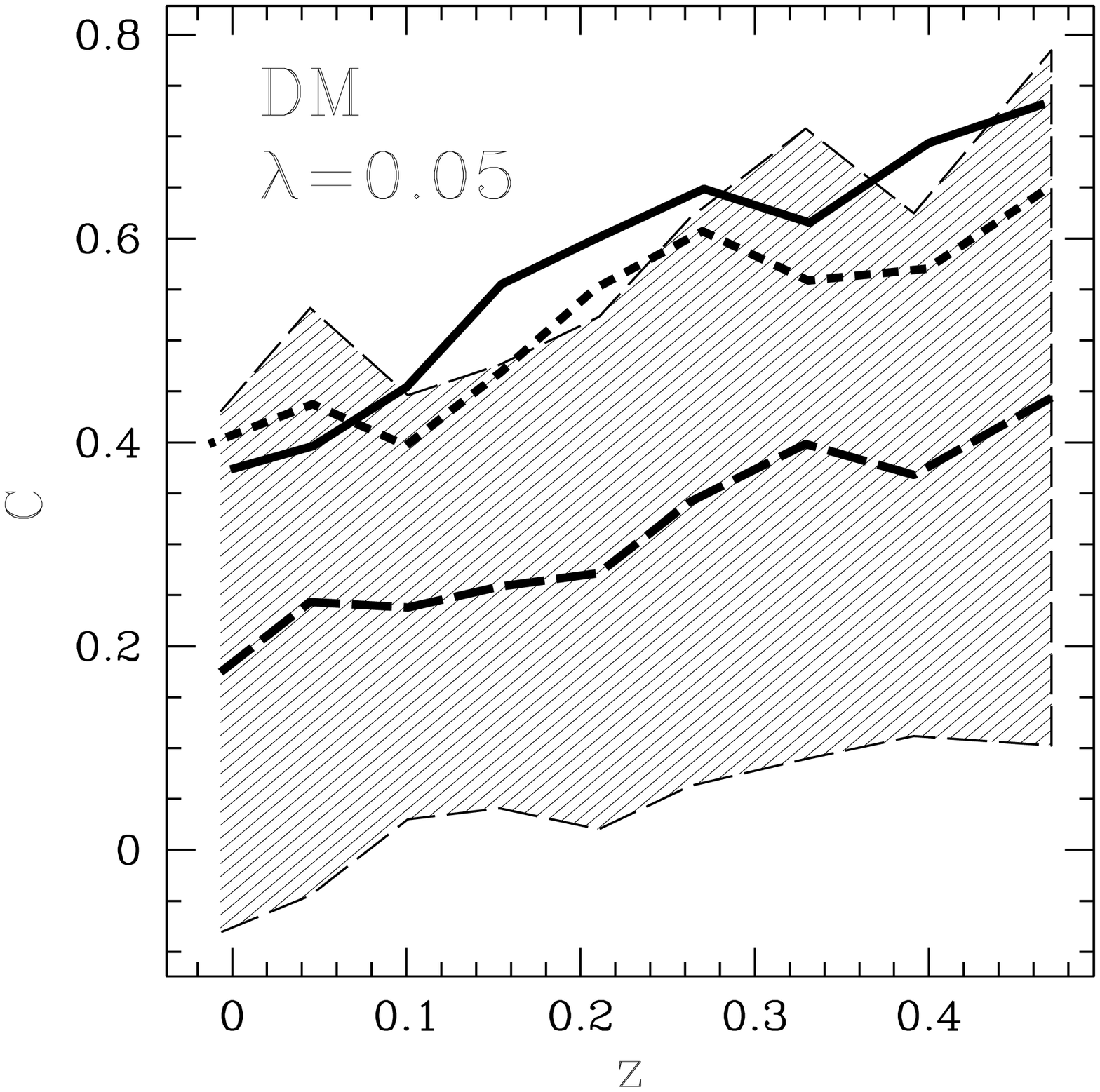}
\includegraphics[height=4.9cm]{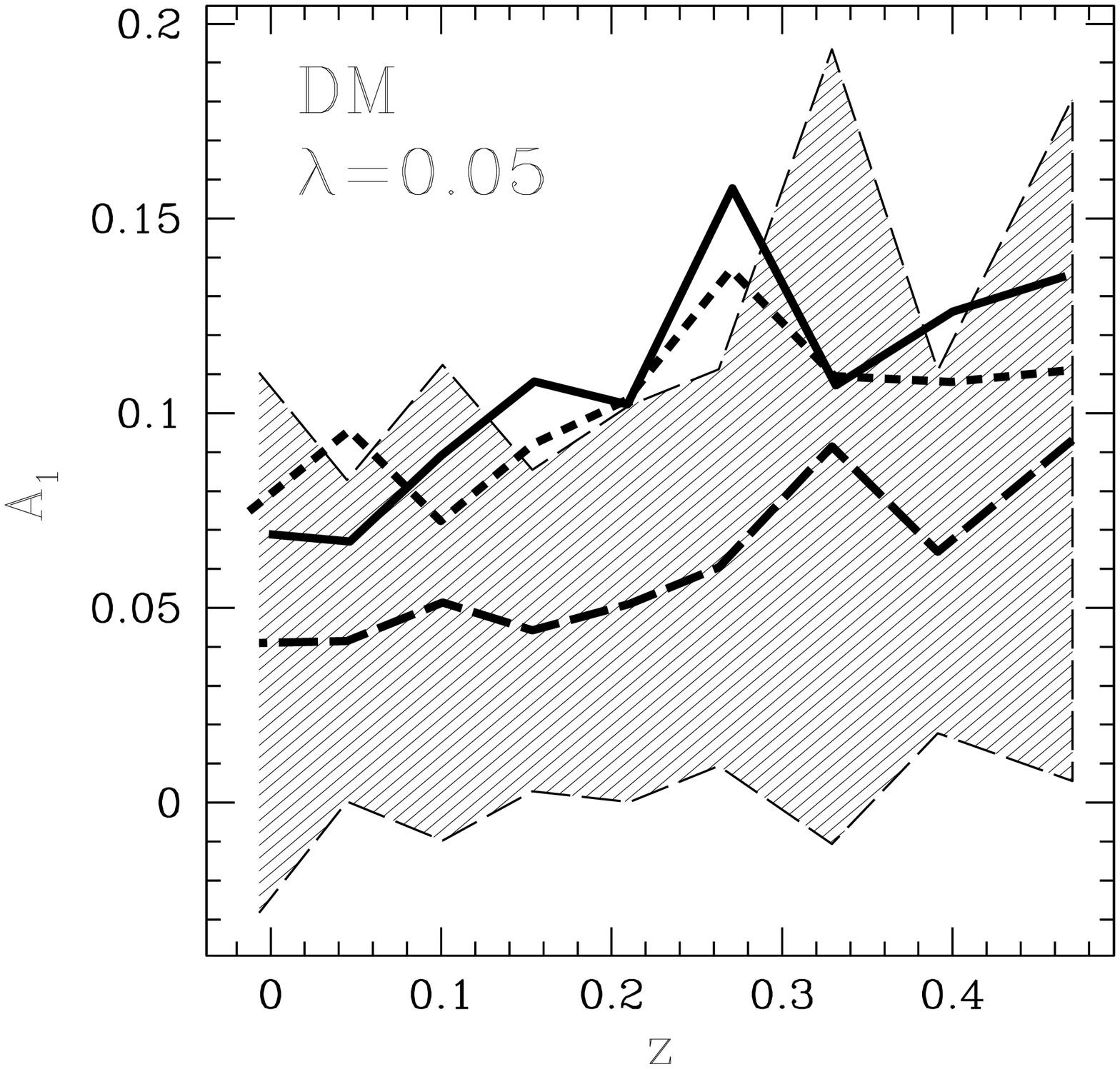}
\includegraphics[height=4.9cm]{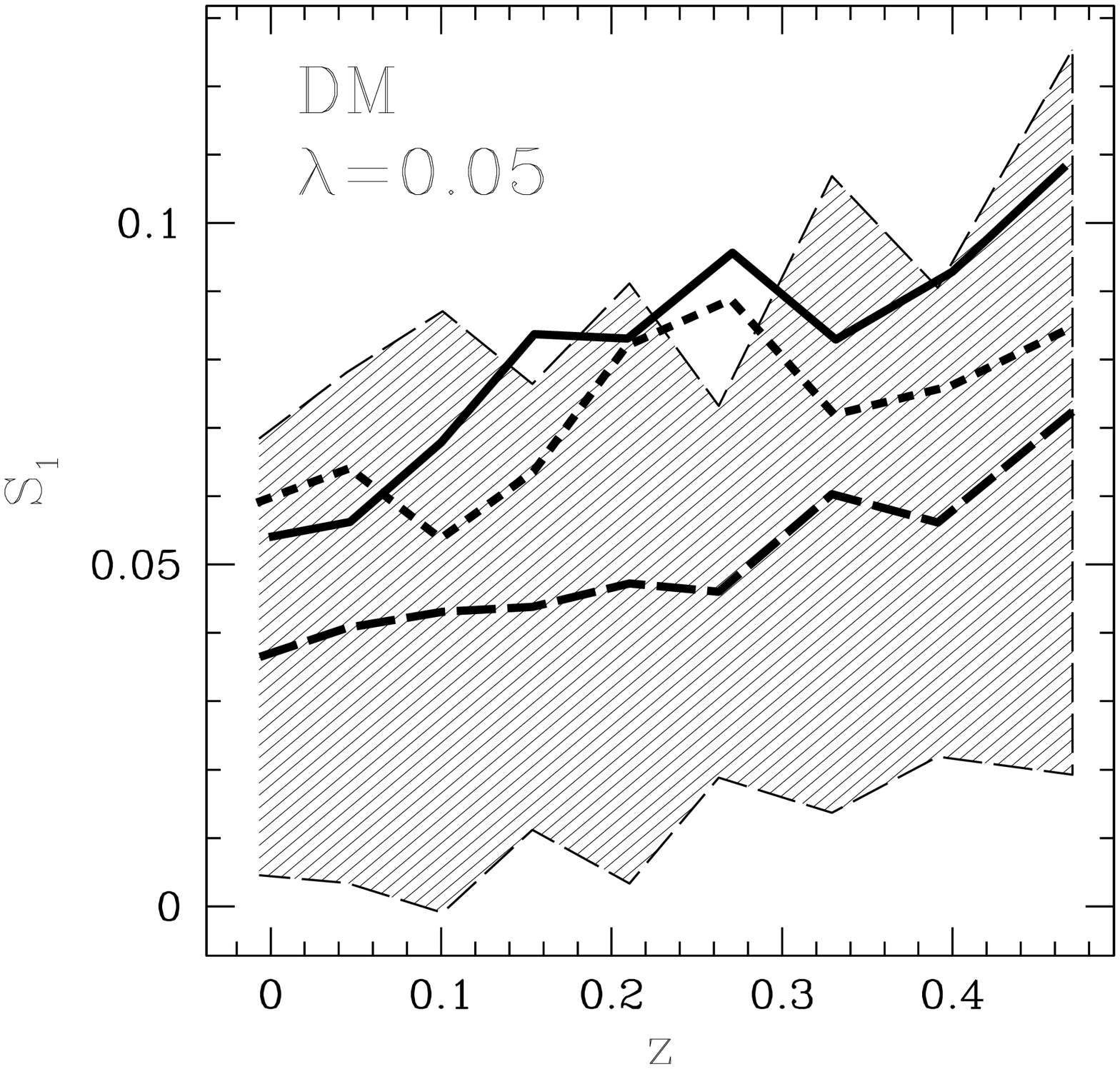}
\caption{   The  morphological   evolution  of   the  DM   within  the
clusters. The ensemble-averaged  structure functions \emph{clumpiness},
\emph{shape and asymmetry}, and \emph{shift of morphological properties}
for  a  smoothing  length  of  $0.05\mpch$ are  shown  vs.   redshift.
\models\; Contrary to Fig.~\ref{fig:compg0.05},  here we probe the DM.
We consider  a spherical window  of radius $r_{\rm w}=1.4  \mpch$; for
the          units         see         the          caption         of
Fig.~\ref{fig:compg0.05}.\label{fig:compd0.05}   This  time,  we
also  show the  one-sigma fluctuations  around the  average structure
functions for the \ldm model.  Note, that large fluctuations are to be
expected on  physical grounds  because of the  wide range  of clusters
investigated,  having different formation  times and  environments. We
conclude that  one has to  consider the whole  morphology distribution
within the  samples in order  to get significant claims  regarding the
background cosmology (see below).}
\end{figure*}
\\
So far we concentrated on the morphological evolution as traced by the
X-ray  luminosity  and thus  the  X-ray gas.   But  is  also the  DM
morphology  different for  the  cosmological models?   The results  in
Fig.~\ref{fig:compd0.05}  show the  mean substructure  evolution for
galaxy  clusters ($\lambda  =  0.05\mpch$) and  indicate  that the  DM
morphology  in  clusters  is  even  more sensitive  to  the  cosmological
background than the gas.
\\
In order  to strengthen our claims  and to compare  the performance of
the gas-  and the  DM-morphology in a  systematic way, we  take into
account  the whole  distribution of  the structure  functions  for our
cluster samples.  The Kolmogorov-Smirnow test  is a suitable  tool to
answer the  question whether two data  samples are likely  to be drawn
from  the same  distribution.  It  measures the  distance  between two
cumulative distributions  $D_1(X) = p(x<X)$ and  $D_2$ (estimated from
the empirical data) via
\begin{equation}
d_{\rm KS}  = \underset{x\in {\mathbb{R}}}{ {\rm max} }\{ |D_1 (x) -D_2 (x) |\}\;\;\;.
\end{equation}
For any value of $d_{\rm KS}$ one can estimate the probability $p_{\rm
KS}$  that  the  same  random  process  generates  two  samples  being
``farther away'' from  each other than $d_{\rm KS}$.   If $d_{\rm KS}$
is large  and, accordingly, $p_{\rm  KS}$ is small,  the distributions
under investigation are likely to stem from two different ensembles.
\begin{table*} 
\centering    
\caption{The discriminative power of the structure functions regarding
the   cosmological  background  models.    The  X-ray   morphology  is
considered  for a  smoothing length  of $\lambda  = 0.05\mpch$  and at
redshift zero.  For each pair  of background models, we  calculate the
Kolmogorov-Smirnow distance $d_{\rm  KS}$ between the distributions of
the   structure  functions  as   well  as   the  probability   of  the
nullhypothesis.   As  one  can   see,  the  gas  substructure  clearly
discriminates     between    the     high-    and     the    low-$\om$
models. \label{tab:ks-test}}
\begin{tabular}{lllllll} 
\multicolumn{7}{c}{ Gas, $r_{\rm w} = 1.40\mpch,\, z = 0,\,\lambda = 0.05\mpch$  }  \\\hline     & \multicolumn{2}{c}{CDM -- $\Lambda$CDM }  & \multicolumn{2}{c}{CHDM -- $\Lambda$CDM }  & \multicolumn{2}{c}{CDM -- CHDM} \\ \hline
       &  $d_{\rm KS}$&  $p_{\rm KS}$  &  $d_{\rm KS}$&  $p_{\rm KS}$  &  $d_{\rm KS}$&  $p_{\rm KS}$ \\ \hline
 C      &\; 0.29\;  &\;  7.7 \%\;    &\; 0.31\;  &\;  4.5 \%\;    &\; 0.17\;  &\;  57.7 \%\;    \\ \hline 
 A$_1$  &\; 0.35\;  &\;  1.6 \%\;    &\; 0.40\;  &\;  0.4 \%\;    &\; 0.12\;  &\;  91.4 \%\;    \\ \hline 
 S$_1$  &\; 0.36\;  &\;  1.2 \%\;    &\; 0.38\;  &\;  0.7 \%\;    &\; 0.15\;  &\;  71.8 \%\;    
\end{tabular} 
\end{table*} 
Table~\ref{tab:ks-test}  presents results  of a  KS test  for redshift
$z=0$ and a smoothing length of $\lambda=0.05\mpch$.  The small values
of $p_{\rm KS}$  show that the distinction of  the cosmological models
by means of the structure  functions is effective regarding the values
of  the cosmological  parameters. A  systematic imprint  of  the power
spectrum  on  cluster  morphology,   however,  does  not  seem  to  be
noticeable.  The DM  is even  better  than the  gas in  discriminating
between the background cosmologies. For instance, in comparing the CDM
and the \ldm model, the  probability of the null hypothesis is smaller
than $10^{-5}$ using the clumpiness ($\lambda=0.1\mpch$).
%
%
%
\\
So far, we  investigated only one smoothing scale and  one size of the
cluster  window;  but one  may  ask  which  cluster regions  are  most
interesting for  a distinction  between cosmological models  and which
smoothing lengths  are optimal  for our purposes.  In order  to answer
these questions, we first focus  on the gas and estimate our structure
functions for  all cluster samples at  redshift $z=0$ for  a number of
smoothing lengths  and scales of  the cluster window. For  each window
and smoothing  scale we calculate  the KS-distances between  our three
models.   The results  show that  for  the clumpiness
small  smoothing lengths  $\lambda$  are more  favourable than  larger
ones.  For  lower resolutions,  therefore, the subclumps  are smeared
out, and the clumpiness is dominated by random fluctuations.
\\
On  the other  hand, the  discriminative  power of  the clumpiness  is
enhanced for larger scales of the window. The reason is that subclumps
which have  not yet merged with  the main cluster component  are to be
found  at the  outer  cluster parts.   The  other structure  functions
mostly depend only relatively weakly on the window scale.  We conclude
that  the outer  cluster regions,  which  probably have  not yet  been
virialised, are  of more interest for the  cosmologist.  Moreover, the
values  of the  structure  functions rise  for  larger windows.   This
confirms  results by~\citet{valdarnini:cluster},  who found  a similar
behaviour  (see, e.g.,  their Table  5).   For the  DM morphology  one
obtains comparable results.
\section{Fundamental plane relations }
\label{sec:fp}
Using cluster simulations one can test the basic assumption behind the
morphology-cosmology  connection presuming  that the  morphology  of a
cluster mirrors  its inner dynamical state reliably.   In this section
we  try to bring  together morphology  and inner  state of  our galaxy
clusters.  Mostly, we focus on the \ldm model as the nowadays favoured
one.
\\  
Observationally there is evidence  that clusters of galaxies undergo a
dynamical  evolution   leading  to   a  sort  of   equilibrium.   This
equilibrium seems to be  manifest in \emph{fundamental plane relations}
holding within  three-dimensional spaces of  global cluster parameters
where clusters tend to populate a plane.  Since the cluster parameters
are logarithms  of observable  quantities, the fundamental  plane (FP)
corresponds  to  a  power   law  constraint  among  the  real  cluster
parameters.   Usually  fundamental plane  relations  are explained  in
terms of the virial theorem of~\citep{chandrasekhar:tensor-vt}, which,
however, is  strictly valid  only for isolated  systems~{}\citep[for a
discussion see][]{fritsch:fp}.
\\
There  are several  interesting spaces  of global  cluster parameters,
depending on  whether optical or  X-ray data are  available.  Usually,
the  scale of  the cluster,  an estimate  of its  mass and  a quantity
related  to its  kinetic energy  like the  velocity dispersion  of the
galaxies or the temperature of the X-ray emitting gas are considered,
see,   e.g.,~\citet{schaeffer:fp,adami:fp}  for   optical  fundamental
planes and \citet{annis:fp,fritsch:fp,fujita:fp1} for X-ray clusters.
In each  case, indirectly,  the potential and  the kinetic  energy are
referred to.
\\  
\citet{fritsch:fp} showed that the substructure of  a cluster is
correlated  to  its distance  from  the  fundamental  plane using  the
COSMOS/APM and the ROSAT data.  In the spirit of their work, we try to
establish a similar connection for simulated X-ray clusters.
\\  
%
\subsection{Fundamental plane relations for the simulated clusters}
\begin{figure*}
\centering
\includegraphics[width=4.7cm]{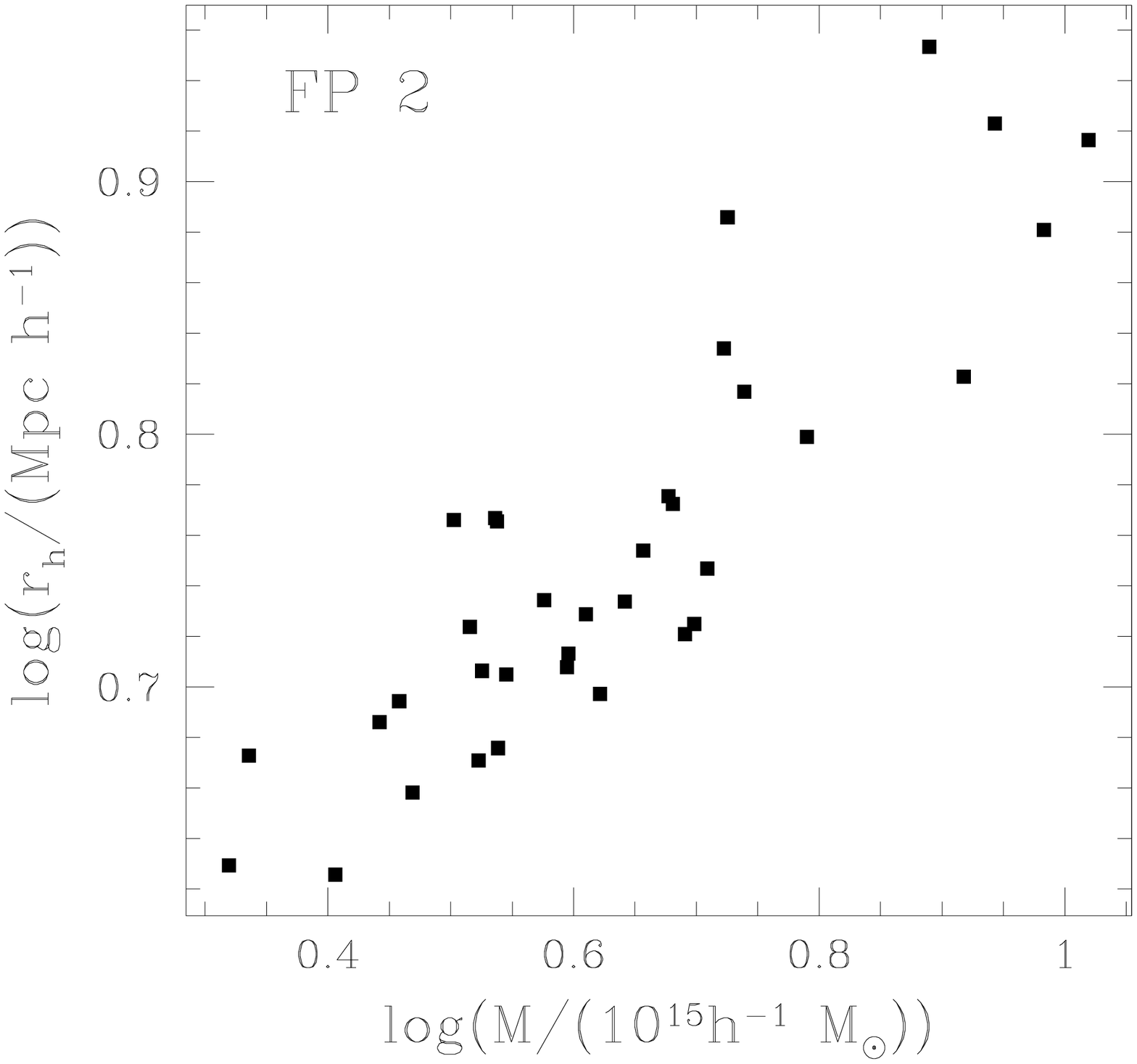}
\includegraphics[width=4.7cm]{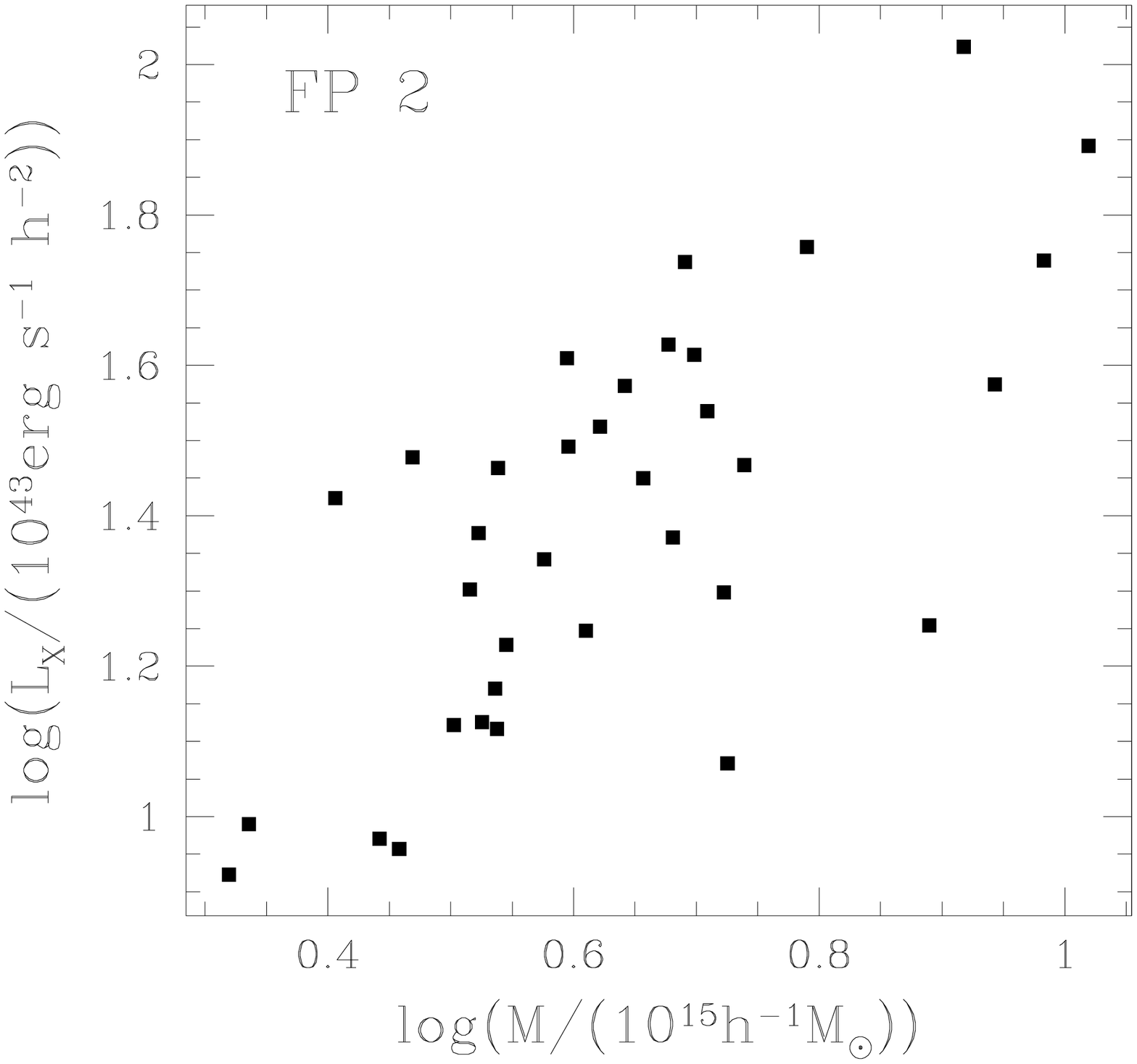}
\includegraphics[width=4.7cm]{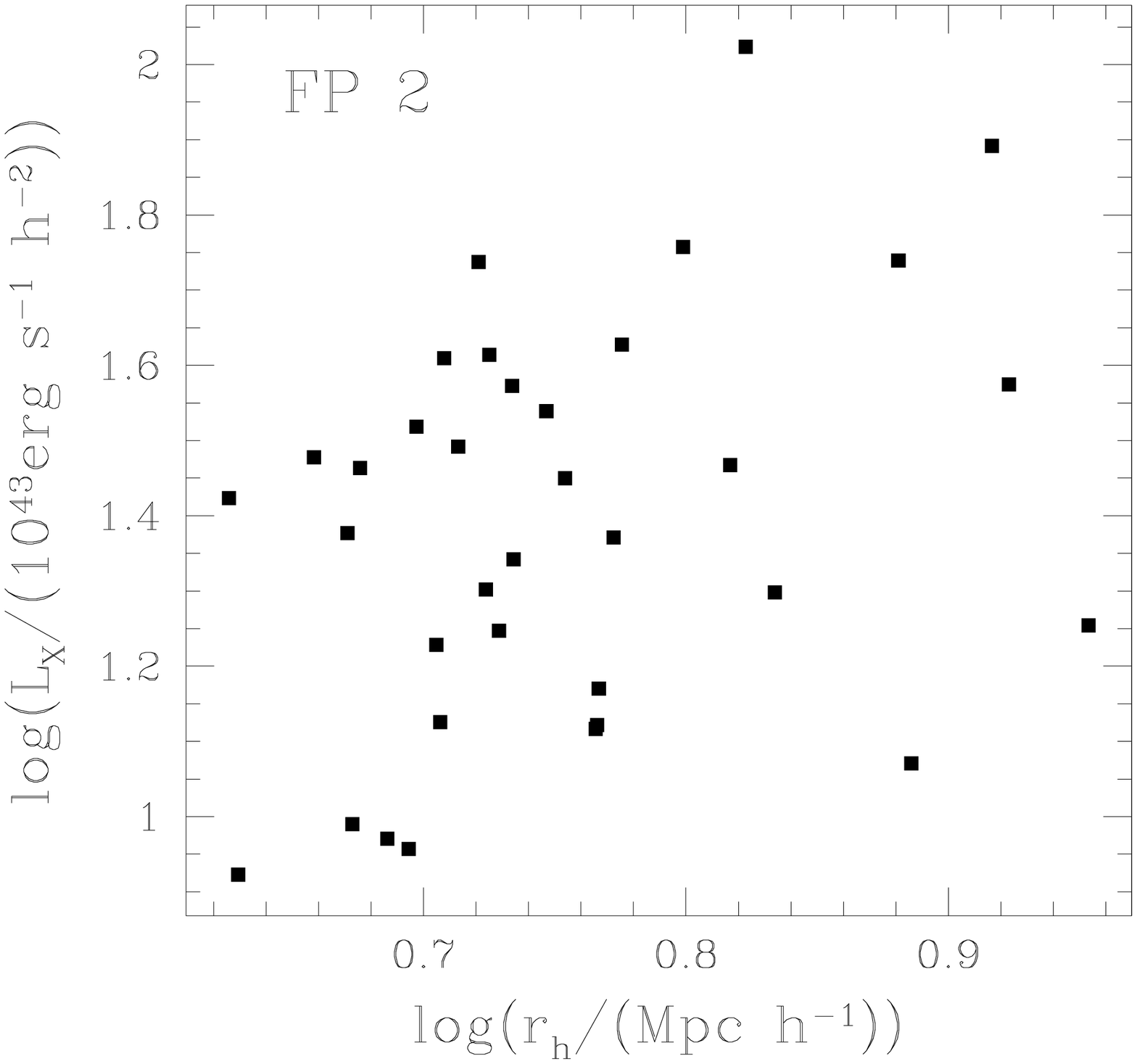}
\includegraphics[width=4.7cm]{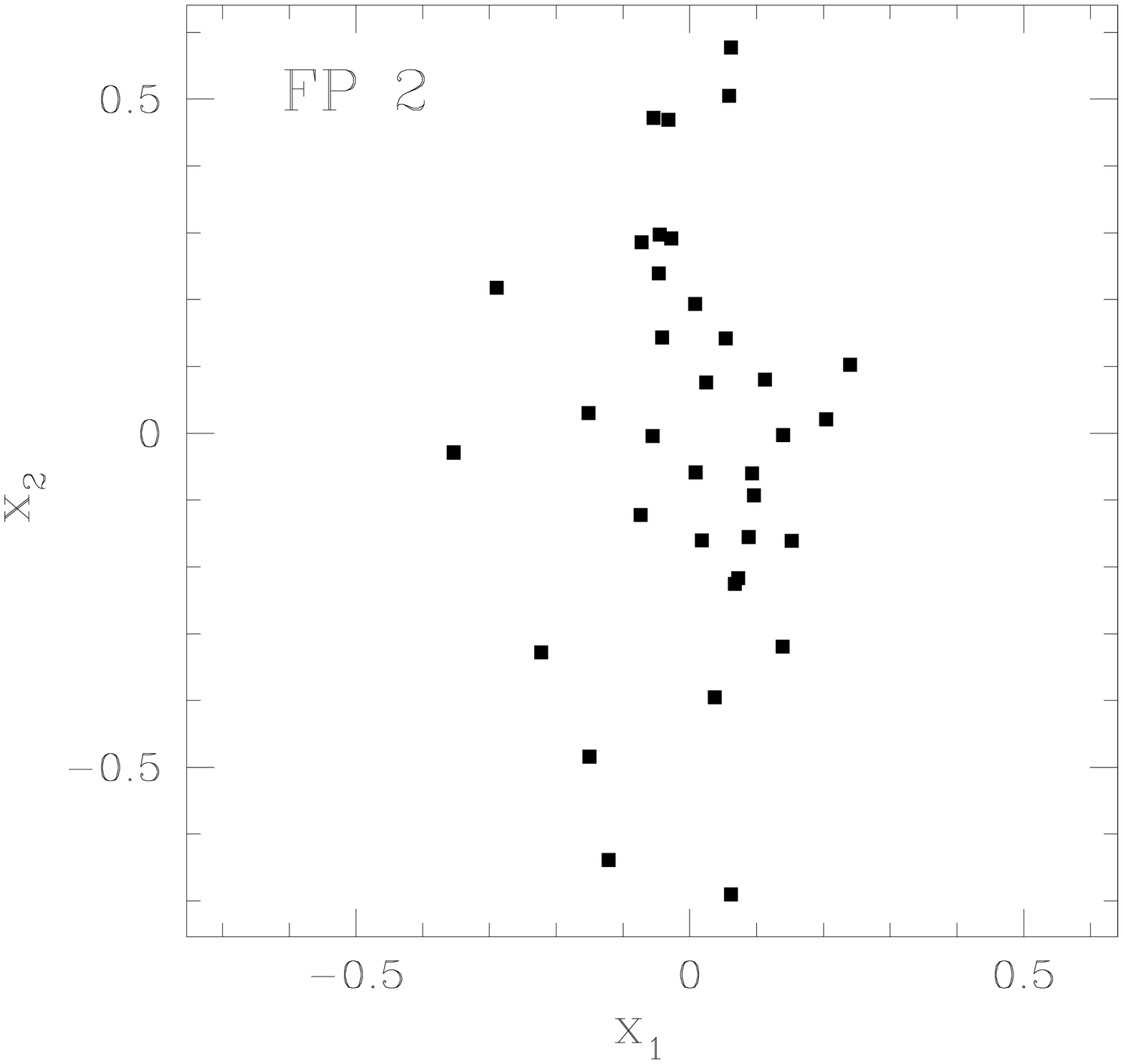}
\includegraphics[width=4.7cm]{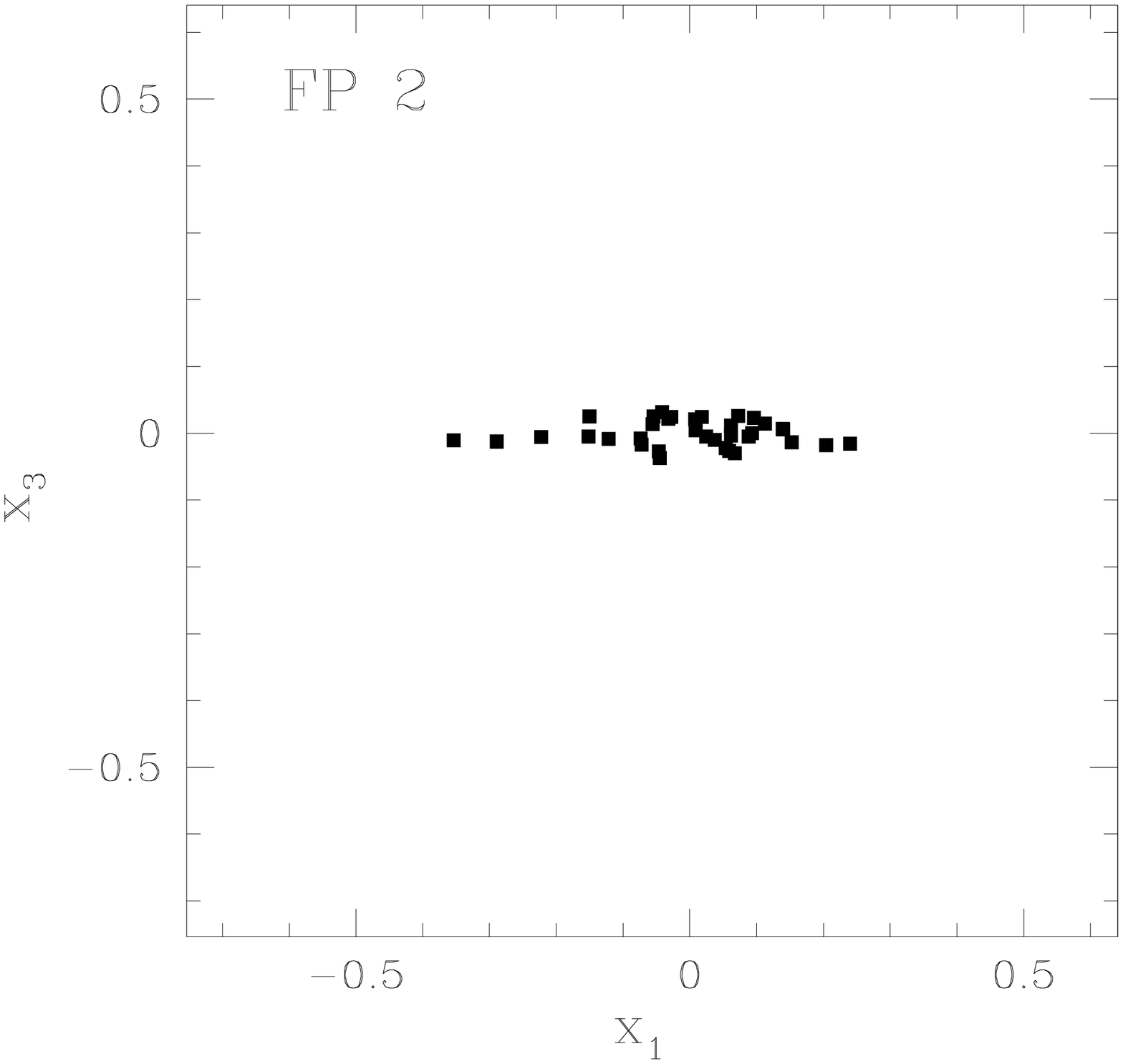}
\includegraphics[width=4.7cm]{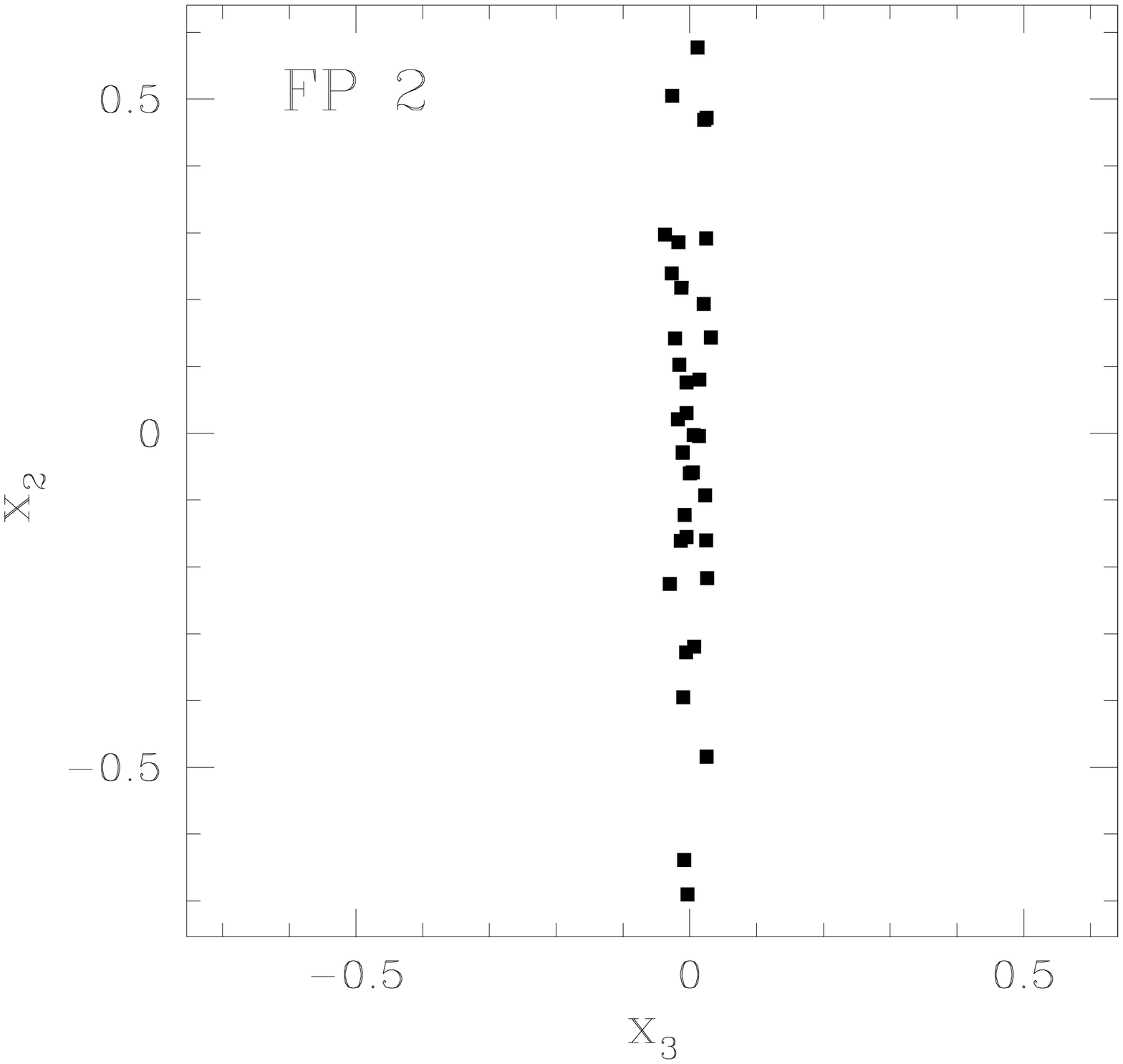}
\caption{ The  fundamental structure in the second  parameter space as
defined  in  Table~\ref{tab:fpd}.   We  consider  the  \ldm  model  at
redshift  zero.   The  first  row  shows the  dependencies  among  the
logarithmised physical parameters for all clusters. In the second row,
we  give  a  visual  impression  of the  fundamental  structure.   The
coordinate axes $x_1$,  $x_2$ and $x_3$ are determined  in such a way,
that the FP coincides with the $x_1-x_2$ plane.
\label{fig:fp_sight}}
\end{figure*}

\begin{table*} 
\centering    
\caption{
Summary of the cluster parameter spaces investigated: the $i$th
parameter space is spanned by the parameters $P_i^{j=0,1,2}$ for $i=1,..,3$.
\label{tab:fpd}
}
\begin{tabular}{l|lll} 
Par. Space $i$ & \multicolumn{3}{c}{Parameters $P_i^j$}  \\  
 No.  &  $j=0$ &  $j=1$ &  $j=2$       \\ \hline 
 $i=1$ & $\log \left(M_{200}/(10^{15}h^{-1} \msolar) \right)$ &
 $\log \left(r_{\rm h}/(100 \kpch)\right)$ & $\log \left( T/(10^7 {\rm \mbox{K}})\right)$    \\ \hline 
 $i=2$ & $  \log \left(M_{200}/(10^{15}h^{-1} \msolar) \right)$ &
 $\log \left(r_{\rm h}/(100 \kpch)\right)$ & $ \log \left( L_{\rm X}/
 (10^{43} {\rm erg s} ^{-1} h^{-2})\right)$    \\ \hline 
 $i=3$ & $ \log \left(M_{200}/(10^{15}h^{-1} \msolar) \right)$ &
 $\log \left(r_{\rm h}/(100 \kpch)\right)$ & $\log \left( \sigma_{\rm v}/(10^2
 {\rm \mbox{km}\; s}^{-1})\right)$    \\ 
\end{tabular} 
\end{table*} 
Using our simulated clusters,  we test three possible parameter spaces
spanned   by:  1.    the  mass,   the  half-light-radius,   and  the
emission-weighted   X-ray    temperature;   2.    the    mass,   the
half-mass-radius,  and the  X-ray  luminosity; 3.   the mass,  the
half-mass-radius,  and the  velocity  dispersion. In  all of  those
global parameter  spaces, one can observe a  \emph{band-like fundamental
structure}.  This thin band may be fitted either by a plane or a line.
In  Table~\ref{tab:fpd} the global  cluster parameters  are summarised
using a compact  notation which we will use from now  on.  
\\ 
The  parameters defining  the different  cluster parameter  spaces are
estimated  from  the  simulations  as  follows: the  cluster  mass  is
quantified via $M_{200}$  contained within an overdensity $\delta_{\rm
c}$  times   the  critical  density   $\rho_{\rm  c}$  :   $M_{200}  =
\frac{4\pi}{3}   \delta_{\rm  c}   \rho_{\rm   c}  r_{200}^3$,   where
$\delta_{\rm   c}   \simeq  178   \cdot   \omo^{-0.45}$   in  a   flat
cosmology~{}\citep{coles:cosmology} and where $r_{200}$ is the size of
this  overdensity.  $r_{200}$ as  well as  $r_{\rm h}$  are determined
from  the  three-dimensional  mass  distribution around  the  density
maximum, $r_{\rm  h}$ is the half-mass radius.   $T_{\rm em}$ denotes
the emission-weighted temperature of  the gas, calculated from the gas
thermal  energy   assuming  an  ideal  gas.    The  bolometric  X-ray
luminosity  is  defined  as  $L_{\rm  X}=\int  \d  V  (\frac{\rho_{\rm
gas}}{\mu m_{\rm  p}})^2 \Lambda_{\rm c} $, where  $\rho_{\rm gas}$ is
the gas density, $\mu=0.6$ the  mean molecular weight, $m_{\rm P}$ the
proton mass, and $\Lambda_{\rm c}$  the cooling function.  In order to
perform the  volume integration, the  standard SPH estimator  has been
applied~\citep{navarro:cluster-simulation}, the  summation includes all
particles within the virial  radius $r_{200}$. The velocity dispersion
is estimated from all types of simulation particles.  \\
Before  investigating the relationships  between these  parameters, we
ask whether the distributions  of these parameters are consistent with
each other for the different  cosmological models (in the sense of the
KS  test). We  find consistency  apart from  the luminosity  (which is
higher  on average  for the  \ldm  model) and  the half-light  radius
(clusters seem to be more compact within the CHDM model).
\\  
We  fit a  plane and  a line  to each  cluster sample  separately (one
sample  means  one  cosmological  model  at  one  redshift)  using  an
ortho\-gonal                    distance                   regression,
see~\citet{boggs:stable,boggs:odrpack};  this   technique  treats  all
variables the same  way, i.e.  no parameter is a  priori thought of as
dependent on  the others\footnote{Note, that  we do not  lump together
different models or  redshifts for the fittings. In  part, we used the
ODRPACK  software package  for the  orthogonal  distance regression.}.
The planes are parametrised by
\begin{gather}
i{\rm th\; plane}\tag{FP $i$}\\
P_i^0 + \beta_i^1 P_i^1   
+ \beta^2_i P_i^2  = \beta^3_i\;\;\;, \label{eq:fp1}\notag
\end{gather}
the lines are defined via:
\begin{gather}
i{\rm th\; line}\tag{FL $i$}\\
\begin{split}
 P_i^1 &=  \gamma_i^1 
   P_i^0 + \gamma_i^3 \notag\\ 
 P_i^2 &=  \gamma_i^2 P_i^0 + \gamma_i^4 \;\;\;.\notag \label{eq:fl1}
\end{split}
\end{gather}
We  call   the  best-fitting   planes  and  lines   \emph{fundamental
planes/fundamental lines}, respectively.  \\
We show  one of the fundamental  structures in Fig.~\ref{fig:fp_sight}
at   redshift  $z=0$   for  the   \ldm  model.    To  get   a  clearer
representation,  we  fit  a  second   plane  to  our  data  under  the
constraint,  that  it  be  orthogonal  to the  fundamental  plane,  as
\citet{fujita:fp1} did. A second constraint equation among the cluster
parameters should  force the clusters to  lie on a line  in the global
parameter  space,  this  line  should   lie  (more  or  less)  on  the
intersection of  both planes.  We  define a rotated  coordinate system
$\{x_1,x_2,x_3\}$  in such a  way that  the first  (i.e.  fundamental)
plane  coincides  with  the  $x_1-x_2$  plane,  and  the  best-fitting
orthogonal plane lies within  the $x_2-x_3$ plane.  In this coordinate
system, the scatters around both  planes are easily discernible as the
$x_3$-  and $x_1$-values  for  the clusters.   The  morphology of  the
structure obviously  is more band- than  plane-like confirming results
by \citet{fujita:fp1}, who call the structure they find in a different
parameter space  ``the fundamental band''.  This is also true  for the
other  parameter  spaces.   A  visual inspection  of  the  fundamental
structures  shows  furthermore  that  most  outliers,  which  tend  to
prolongate  the line,  wander towards  the bulk  for  lower redshifts.
Note, that in our analysis statistical outliers are not removed.
\begin{table}
\centering
\begin{tabular}{lll}
FP $i$ & model & $M_{200}\propto \left(P_i^{1}\right)^{\beta^1_i} \left(P_i^{2}\right)^{\beta_i^2}$ \\\hline
1 &   CDM  & $M_{200}\propto r_h^{1.68\pm 0.39}T_{\rm em}^{  0.71\pm 0.18  }$  \\
1 &   \ldm & $  M_{200}\propto r_h^{1.62\pm 0.24 }T_{\rm em}^{  0.56\pm 0.15 }$  \\
1 &   CHDM & $  M_{200}\propto r_h^{1.66\pm 0.27}T_{\rm em}^{   0.69\pm 0.13 }$  \\
2 &   CDM  & $  M_{200}\propto r_h^{1.62 \pm 0.32  }L_{\rm X}^{  0.31\pm 0.07  }$ \\
2 &   \ldm & $  M_{200}\propto r_h^{1.60\pm 0.17 }L_{\rm X}^{ 0.29\pm 0.05  }$  \\
2 &   CHDM & $  M_{200}\propto r_h^{1.78\pm 0.26 }L_{\rm X}^{ 0.31\pm 0.06  }$ \\
3 &   CDM  & $  M_{200}\propto r_h^{0.61\pm 0.29 }\sigma^{ 1.96\pm  0.26  }$  \\
3 &   \ldm & $  M_{200}\propto r_h^{1.14\pm 0.38 }\sigma^{  1.90\pm 0.60  }$ \\
3 &   CHDM & $  M_{200}\propto r_h^{0.45\pm 0.43 }\sigma^{  1.83\pm 0.37  }$  \\
\end{tabular}
\caption{ The  best-fitting  fundamental  planes  together  with  the
$95$\%-confidence  regions for  all models  and all  parameter spaces
considered at redshift zero.\label{tab:fp_par_value} }
\end{table}
\\
The   values    of   the   best-fit   parameters    are   listed   in
Table~\ref{tab:fp_par_value}  together  with their  $95\%$-confidence
intervals.   Since  we  have  no  measurement errors  for  the  global
parameters, we can  give error bars only assuming  the goodness of the
fit.  In  order to  probe whether our  fundamental planes  may explain
observed fundamental  plane relations, we compare  the parameters with
simple theoretical  scaling laws and observed  parameters. The problem
about  such  comparisons, however,  is  that  the  definitions of  the
cluster  parameters significantly  depend  on the  techniques used  to
determine  them   from  observations.   Therefore,  we   have  to  use
additional assumptions;  we constrain ourselves  to X-ray fundamental
planes  at redshift zero.   -- From  a theoretical  point of  view the
third  fundamental plane is  the simplest  one.  A  virial equilibrium
requires that $M_{vir}\propto R \sigma^2$  for the whole mass $M$, the
scale $R$ and the velocity dispersion $\sigma^2$ of a cluster being in
virial equilibrium.   If we simply identify these  parameters with the
quantities spanning the third cluster parameter space, we see that our
values for $\beta_3^2$ are  consistent with the virial equilibrium for
all  models;  moreover,  $\beta_3^1$  is compatible  with  the  virial
prediction for  the \ldm model,  and marginally consistent  within the
CDM model, but inconsistent for  the CHDM model. A physical reason may
be that, because  of the high value of $\om$,  a virial equilibrium is
not yet reached for most clusters within the CHDM model.  Perhaps also
the plane-fit is determined by a few clusters not yet in equilibrium;
but certainly larger cluster samples  are required in order to clarify
this point definitely.  \\ The first fundamental plane can be compared
to the results  by~\cite{fujita:fp1}.  Using data from~\cite{mohr:icm}
they find that the central gas density $\rho_{g,0} \propto R_1^{-1.39}
T_{\rm  em}^{1.29}$, where  $R_1$ is  the  core radius.   In order  to
relate  our   parameters  to   theirs  we  estimate   $\rho_{g,0}$  by
$\rho_{g,0} \propto M_{gas}/r_{h}^3 \propto f M/r_{h}^3$, where $M$ is
the whole  mass of  the cluster and  $f$ denotes the  baryon fraction.
Assuming furthermore that $R_1\propto  r_h$ and $M_{200}\propto M$, we
derive from  our first fundamental  plane-fit (for the \ldm  model at
redshift zero; we assume that the additional scaling relations used do
not introduce additional uncertainties)
\begin{equation}
f M_{200} \propto r_h^{1.61\pm 0.24} T^{1.29\pm 0.15} \;\;\;.
\end{equation}
This  result  is  consistent  with  theirs provided  that  the  baryon
fraction  depends relatively  strongly on  the  temperature: $f\propto
T^{0.73}$.   For  comparison, using  the  same  observational data  as
\citet{fujita:fp1},  \citet{mohr:icm} find $f_{ICM}  \propto T^{0.34\pm
0.22}$   within  $r_{500}$,   see  \citet{mohr:icm}   for   the  exact
definitions   of   the   quantities   they   use\footnote{Note,   that
\citet{mohr:icm} use  $90$ \% confidence  regions instead our  $95$ \%
confidence levels.}. For  the other cosmological  models, the
agreement  is better. -- From a  theoretical  point of  view, one  would
expect that for a hydrostatic and a virial equilibrium $M\propto R T$.
For our data the $T$-dependence is slightly weaker, whereas $R$ has a
stronger influence on the mass.
\\ 
For the second parameter  space, we can use
results   by  \citet{fritsch:diss}   which  constitute   the   base  of
\cite{fritsch:fp}.  Assuming that the mass-to-light ratio is
constant for galaxy clusters without scatter, our second fundamental
plane translates into 
\begin{equation}
L_{\rm o}\propto r_h^{1.60\pm 0.17}L_{\rm X}^{ 0.29\pm 0.05  }\;\;\;,
\end{equation}
where  $L_{\rm o}$  is the  optical  luminosity.  On  the other  hand,
putting together  the virial mass  estimate and the  fundamental plane
from \citet{fritsch:diss}, relating $L_{\rm  o}$, $L_{\rm X}$, and the
optical half-light-radius $r_{\rm o}$, we get
\begin{equation}
M_{\rm vir} = \propto L_{\rm X}^{0.35} r_{\rm o}^1 \;\;\;.
\end{equation}
Especially the dependence on the scale is considerably stronger in our
fundamental plane, but again the discrepancies may be explained by the
fact  that our estimates  of the  cluster mass  and scale  differ from
Fritsch' ones. \\ Altogether, our  results are in rough agreement with
most of the theoretical expectations and the observed scaling laws. In
detail, however, there are some inconsistencies to be found; but these
incompabilities may be  explained either with statistical fluctuations
or  by questioning some  of the  assumptions used  in order  to relate
parameters estimated in different ways.
\\
To analyse  the morphologies of  the fundamental structures  and their
redshift evolutions  quantitatively, we investigate  the mean scatters
around     the     fundamental     planes,     $\sigma_{\rm FP}     \equiv
\sqrt{\frac{1}{N}\sum_{i=1}^{\rm N}  d_i^2}$, and  the  orthogonal planes,
$\sigma_{\rm OP}\equiv\sqrt{\frac{1}{N}\sum_{i=1}^{\rm N} \tilde{d}_i^2}$,  
where we  sum up  the  quadratic distances  of the  $N$
clusters from the fundamental planes, $d_i$, and the orthogonal
\begin{figure*}
\centering
\includegraphics[width=4.9cm]{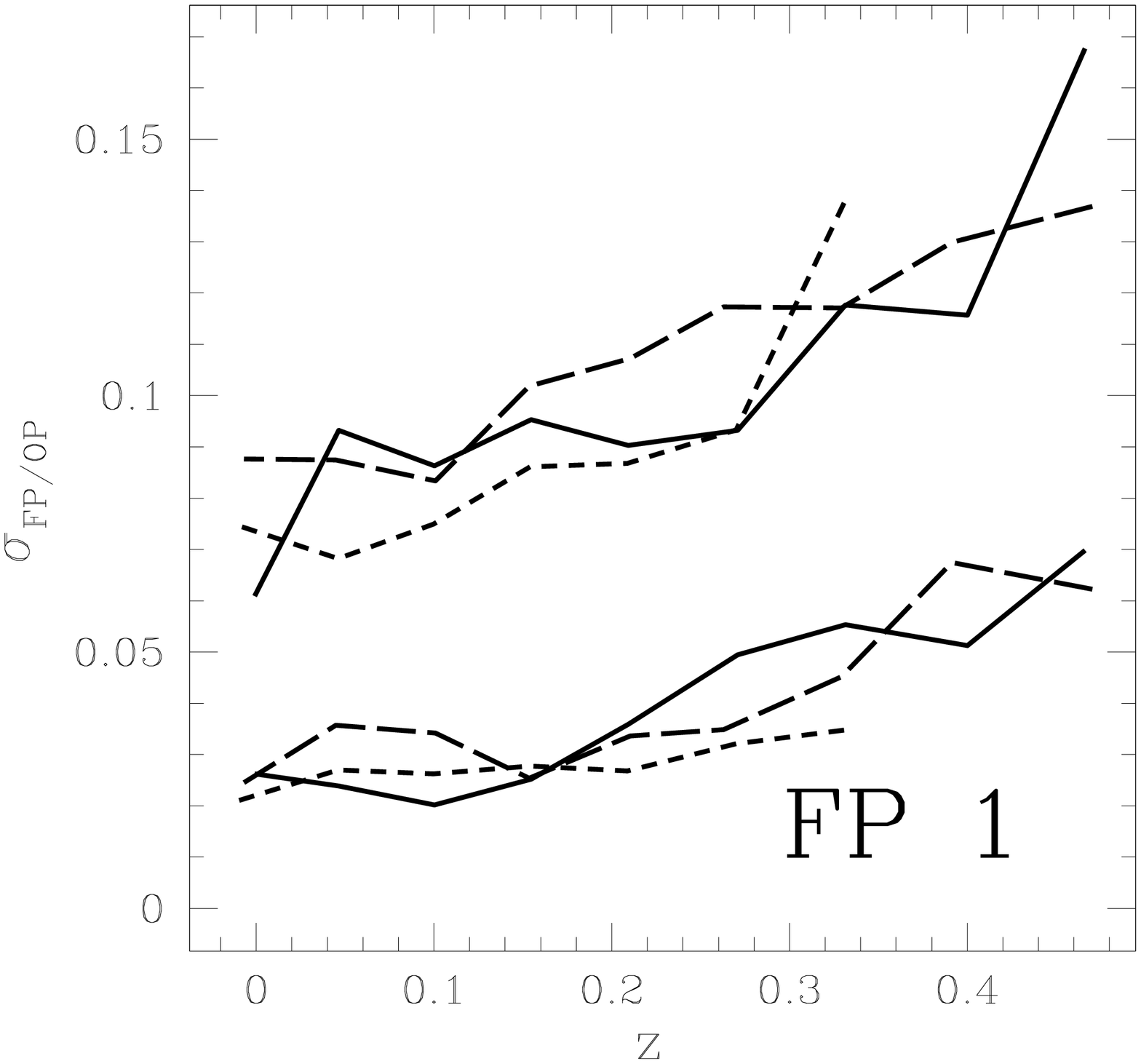}
\includegraphics[width=4.9cm]{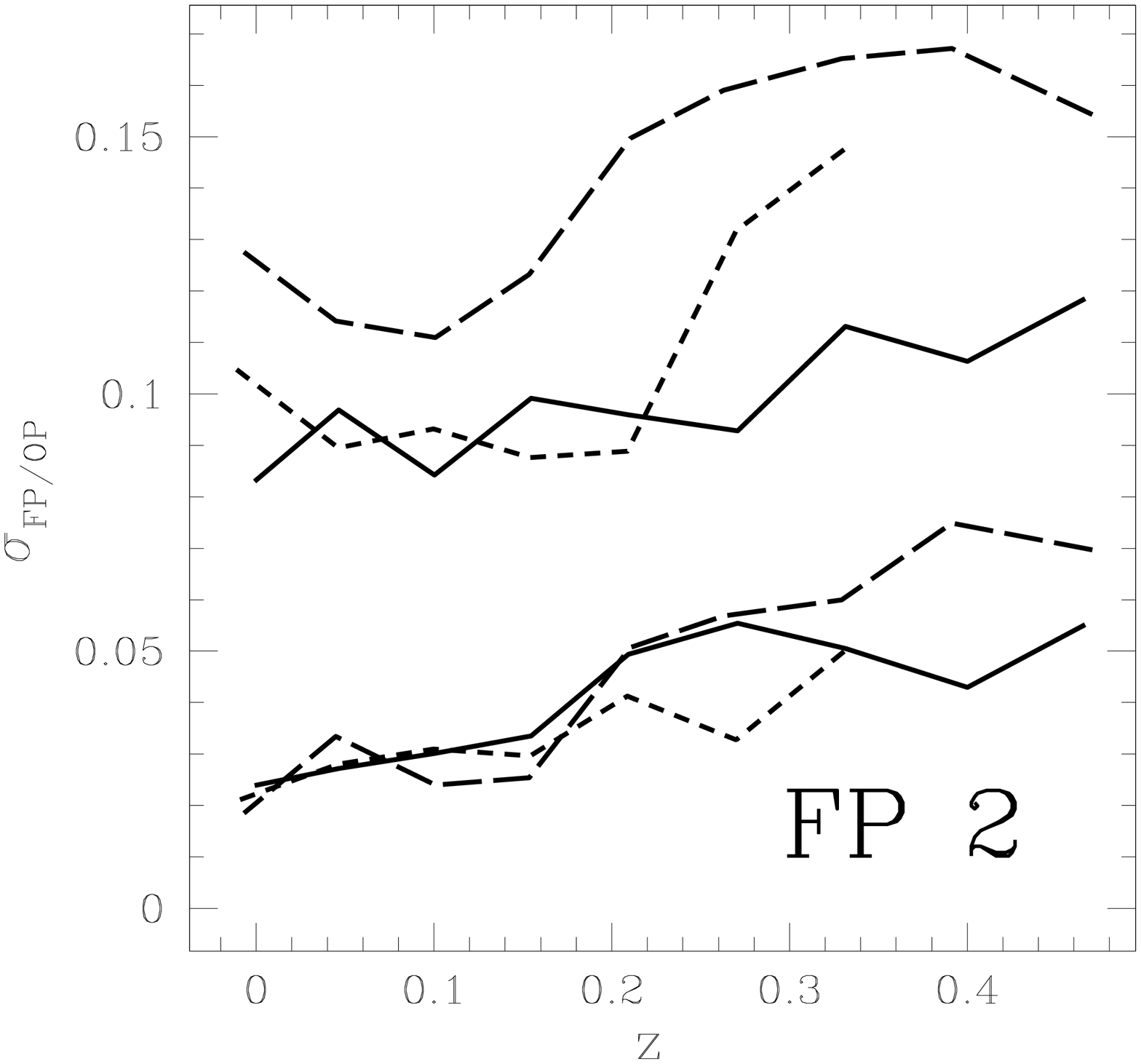}
\includegraphics[width=4.9cm]{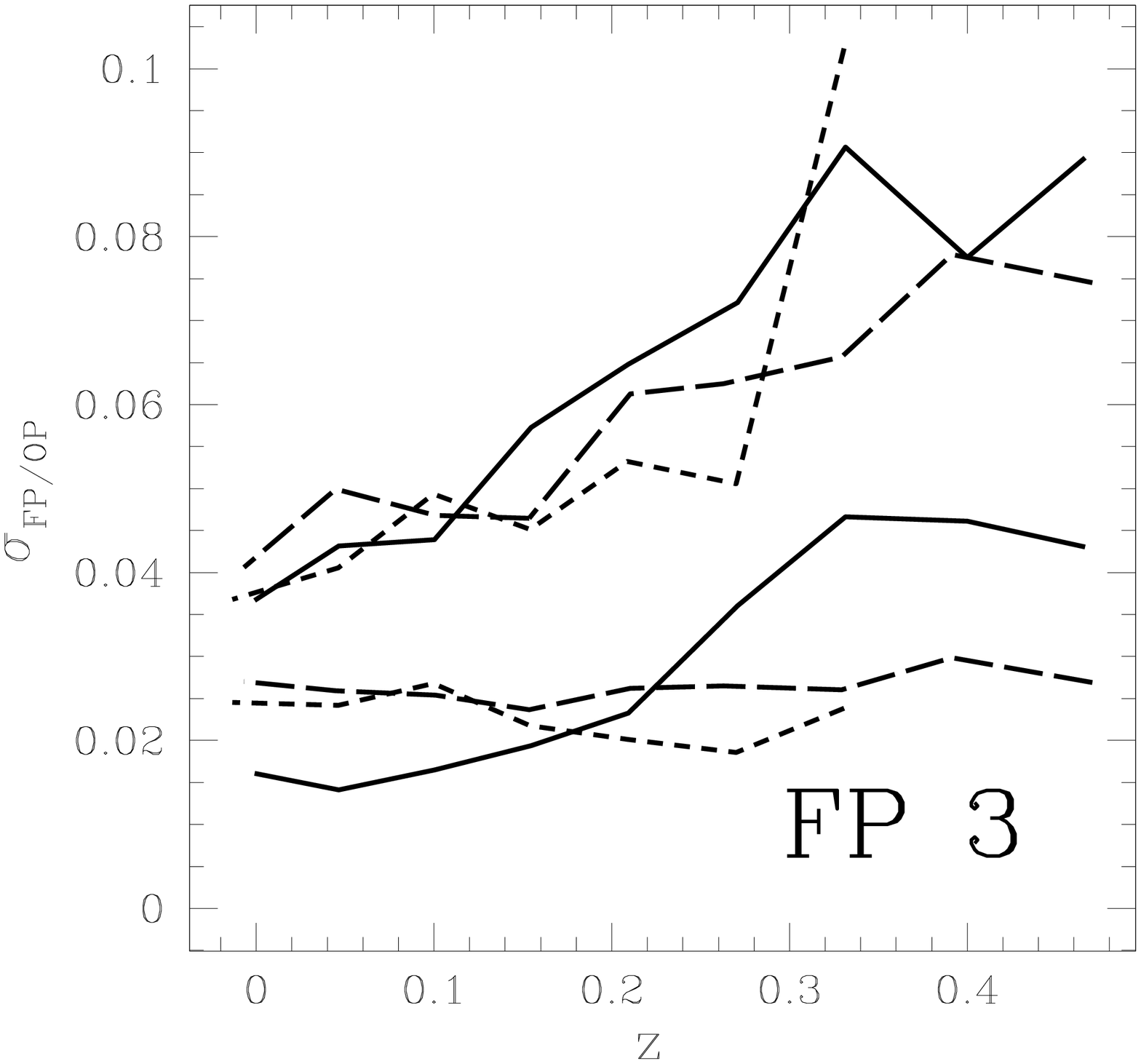}
\caption{For  these  plots  planes  were  fitted  to  the  fundamental
structures in each  parameter space. We show the  mean scatters around
these  fundamental   planes  (lower  curves)  and around  the  best-fitting
orthogonal  planes  (upper  curves)  for  each parameter  space  as  a
function  of  redshift.   Again,  we  have:  \models\;  For  technical
reasons, we consider the CHDM  model starting from $z\sim 0.33$, only.
One sees  that the  scatter around the  fundamental planes  is clearly
smaller than for the orthogonal planes.\label{fig:fp_both_scatter}}
\end{figure*}
\begin{figure*}
\centering
\includegraphics[width=4.9cm]{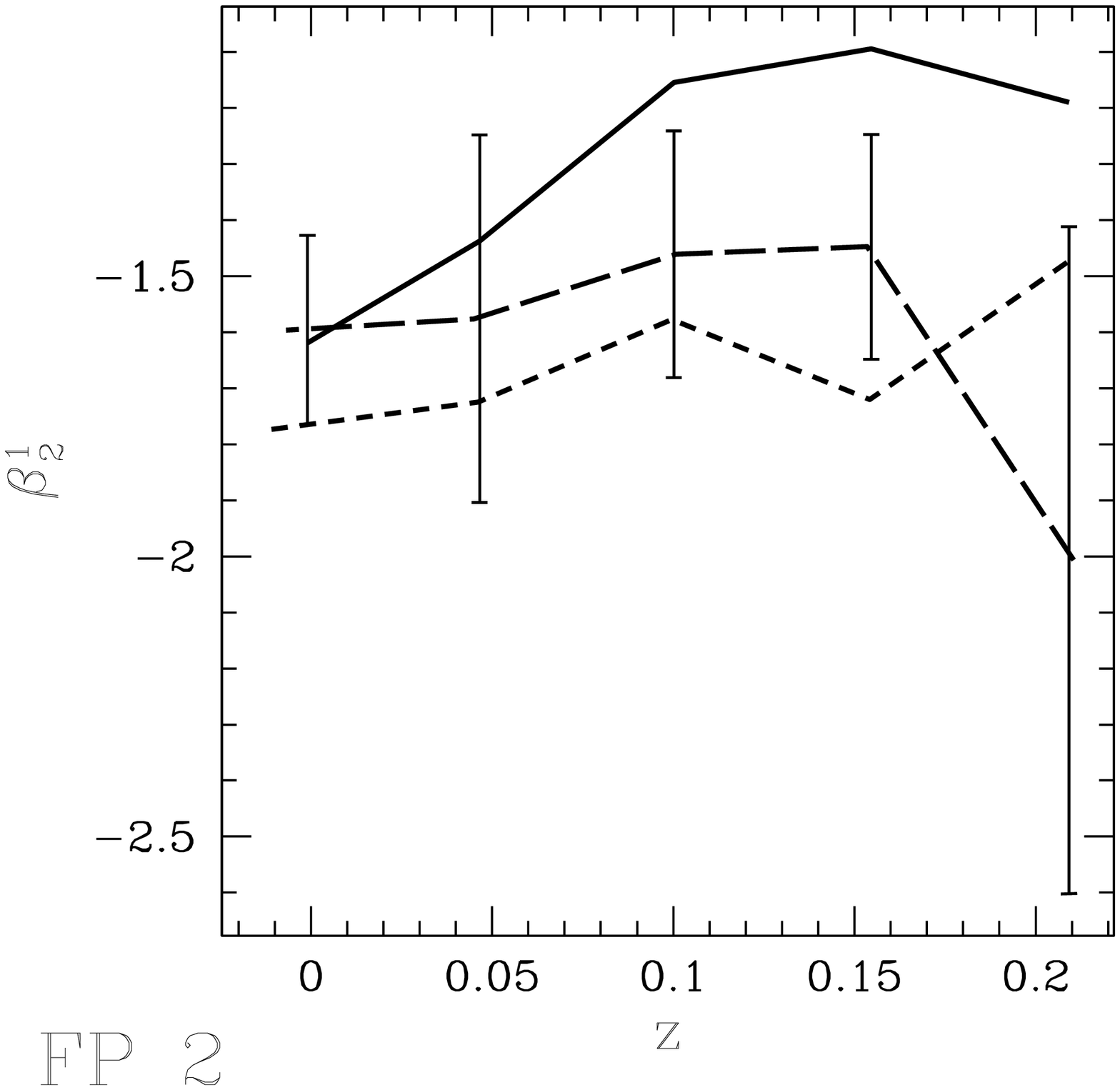}
\includegraphics[width=4.9cm]{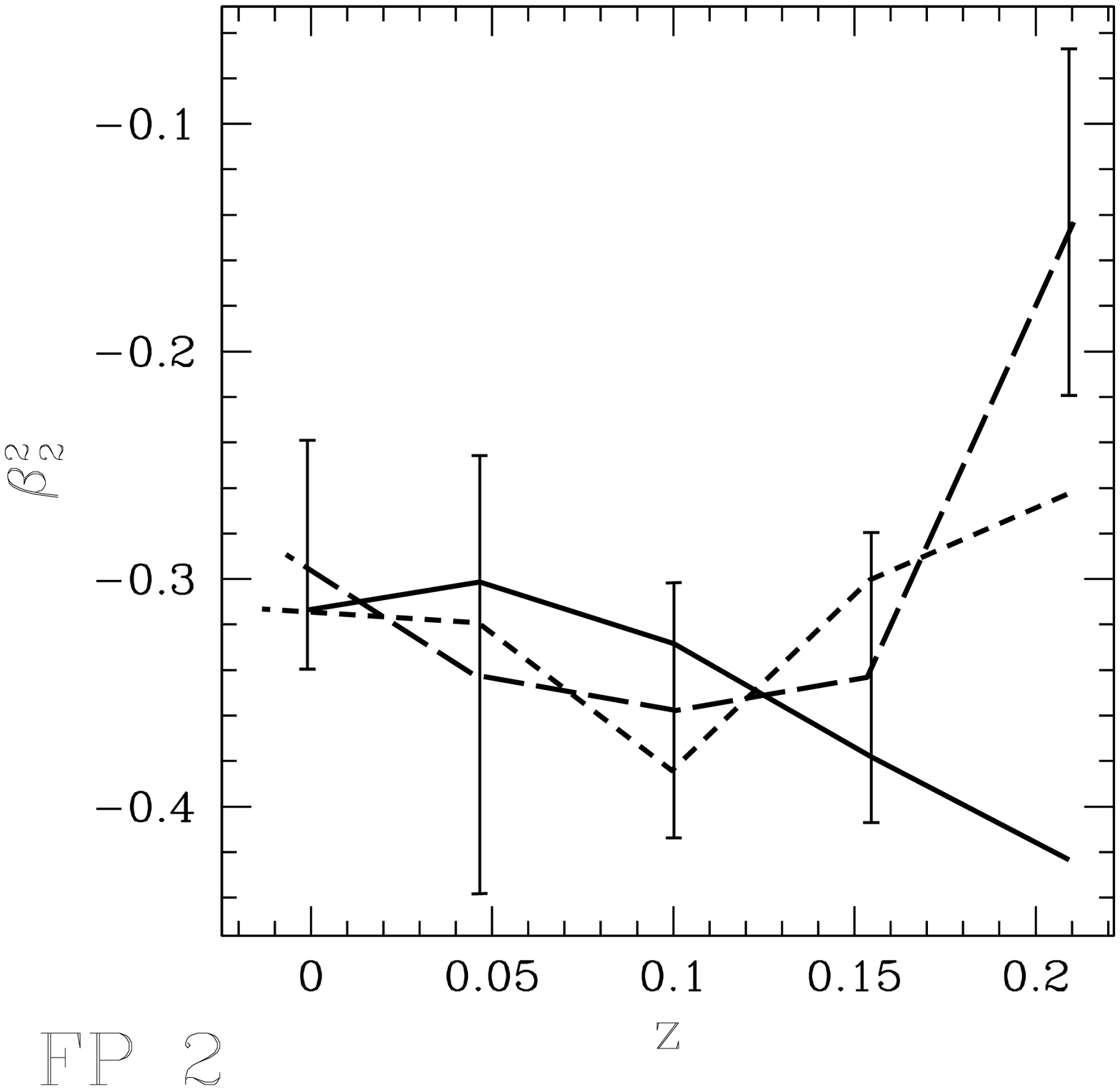}
\includegraphics[width=4.9cm]{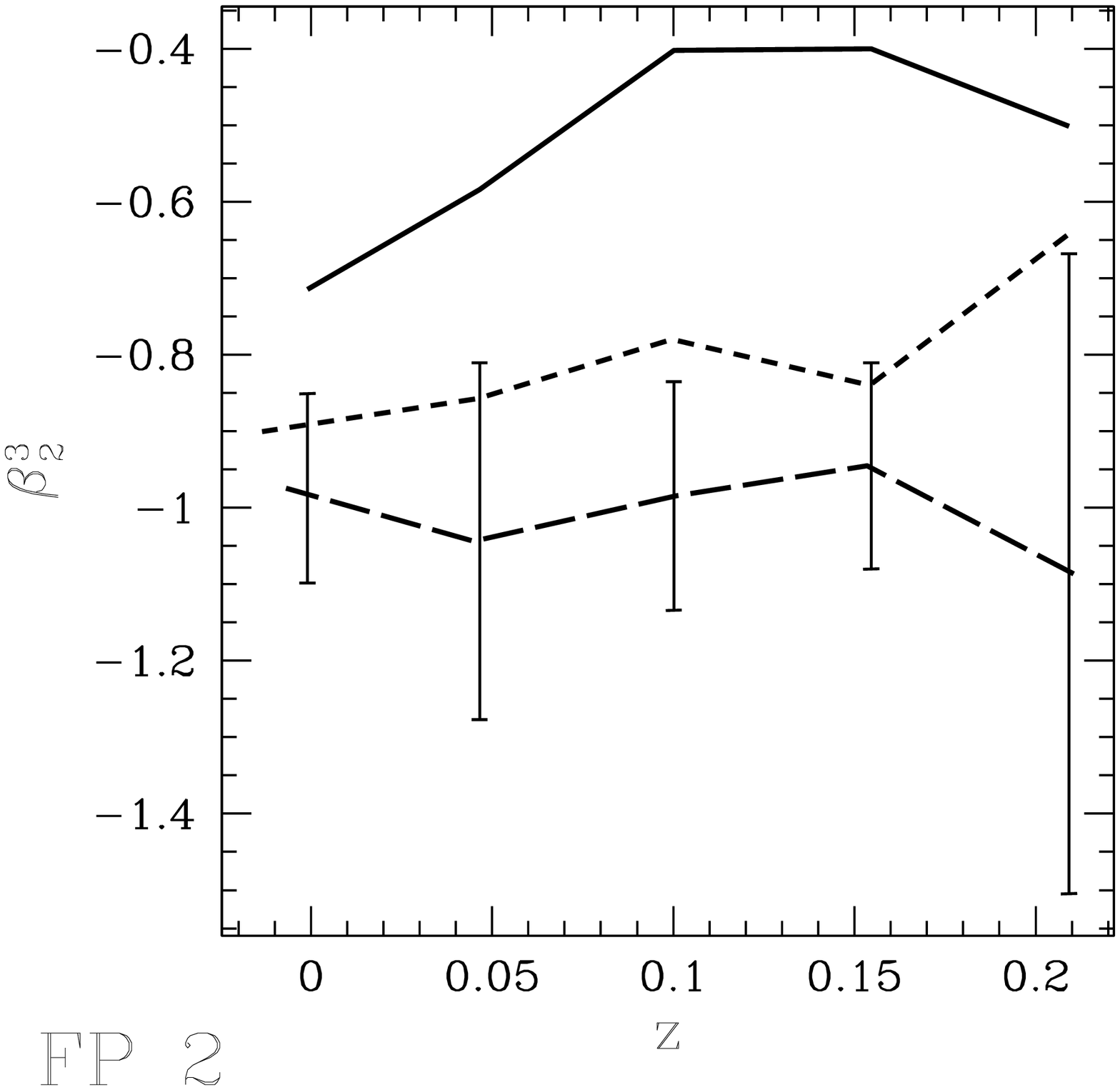}
\caption{The exponents  determining the second fundamental  plane with their
errors at low  redshifts, see Eq.~(FP $i$). Since  we do not have
measurement-like errors,  the 95\%  confidence regions visible  in the
plot    are     estimated    assuming    the     goodness    of    all
fits. \models\label{fig:fp_param_err}}
\end{figure*}
planes,      $\tilde{d}_i$.      As      one     can      see     from
Fig.~\ref{fig:fp_both_scatter}, $\sigma_{\rm FP}$ is  decreasing on
the whole for each of the
first two global parameter spaces. This, however, is not valid for the
third  parameter  space if one fits the fundamental structure using
a plane.   These  details may  indicate,  that  the  band in  the  third
parameter  space is  better fitted  using a  line. This  conclusion is
confirmed if one takes into  account the scatter around the orthogonal
plane, $\sigma_{\rm OP}$:  Fig.~\ref{fig:fp_both_scatter}  shows  that  for  the  third
parameter space, the scatter around the fundamental plane is only
about two times larger than that one around the orthogonal plane.
\\
The   evolution  of   one   set   of  FP   parameters   is  shown   in
Fig.~\ref{fig:fp_param_err}   for  low  redshifts.   The  cosmological
models' confidence intervals, which  were estimated again assuming the
goodness of  the fit, overlap  for small redshifts  ($z\lesssim 0.05$)
indicating the consistency of the models regarding the location of the
second fundamental  plane. Apart from the CDM  model the FP-exponents
do  not  show  any  significant  evolution  for  redshifts  $z\lesssim
0.15$. Similar results hold for the first parameter space.
\\ 
The scatters around the  best-fitting \emph{lines} are decreasing as a
function  of  redshift   in  most  cases  (Fig.~\ref{fig:fl_scatter}).
Especially,  the first  and the  third parameter  space show  a strong
redshift evolution, whereas for the second parameter space results are
less  definitive.   This complements  our  earlier observations,  that
within the  second parameter space  the fundamental structure  is more
plane-like, whereas the third fundamental structure resembles a narrow
band. To  summarise the properties  of the fundamental  structures: in
the first  global parameter  space we see  a band-like  structure, the
structure in the second space can be understood as a plane, whereas in
the  third space  the data  are  better fitted  to a  line. Using  the
corresponding fittings,  the scatters go  down, which we  interpret in
terms of an equilibrium attracting the clusters.
\begin{figure*}
\centering
\includegraphics[width=4.9cm]{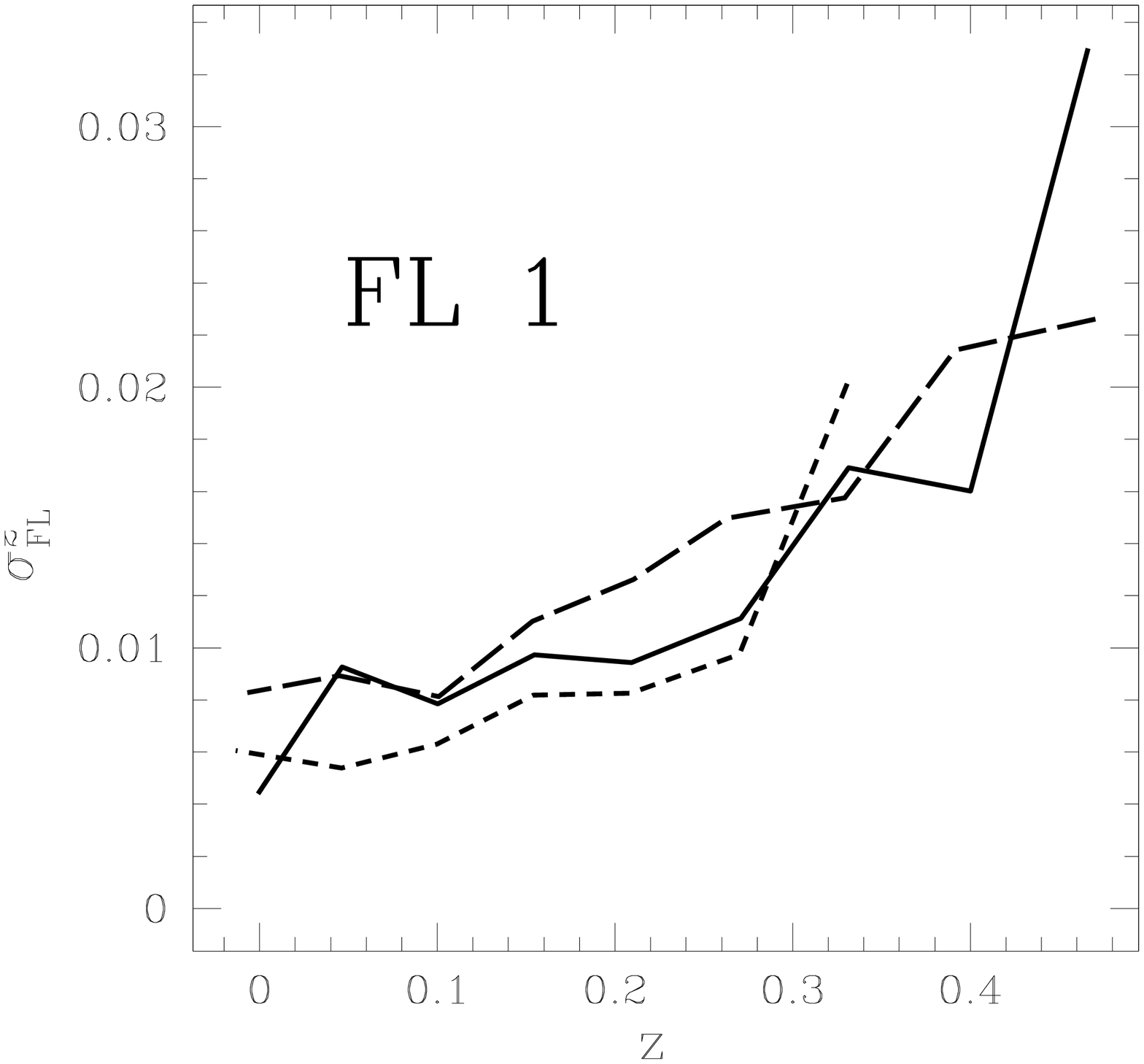}
\includegraphics[width=4.9cm]{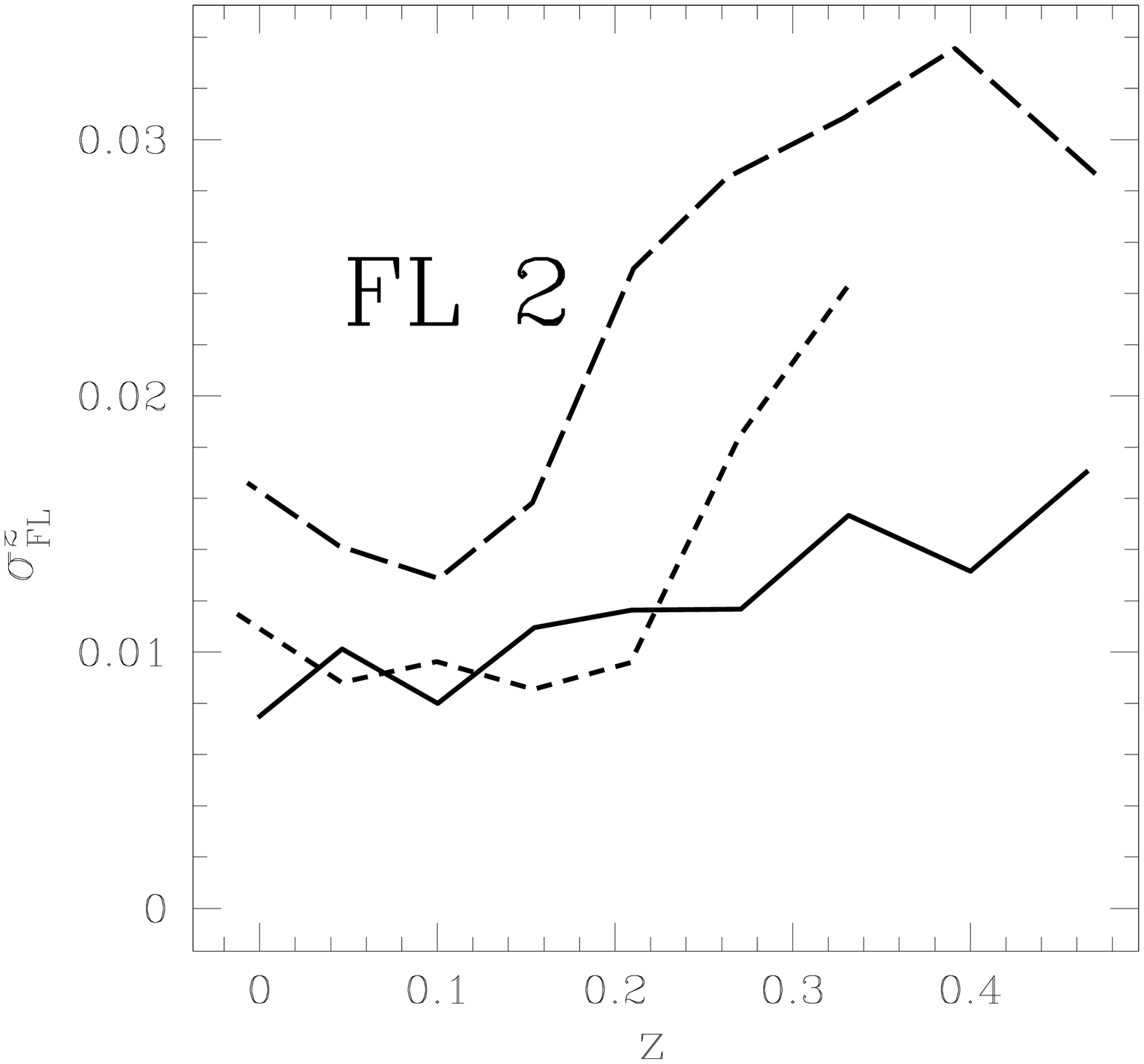}
\includegraphics[width=4.9cm]{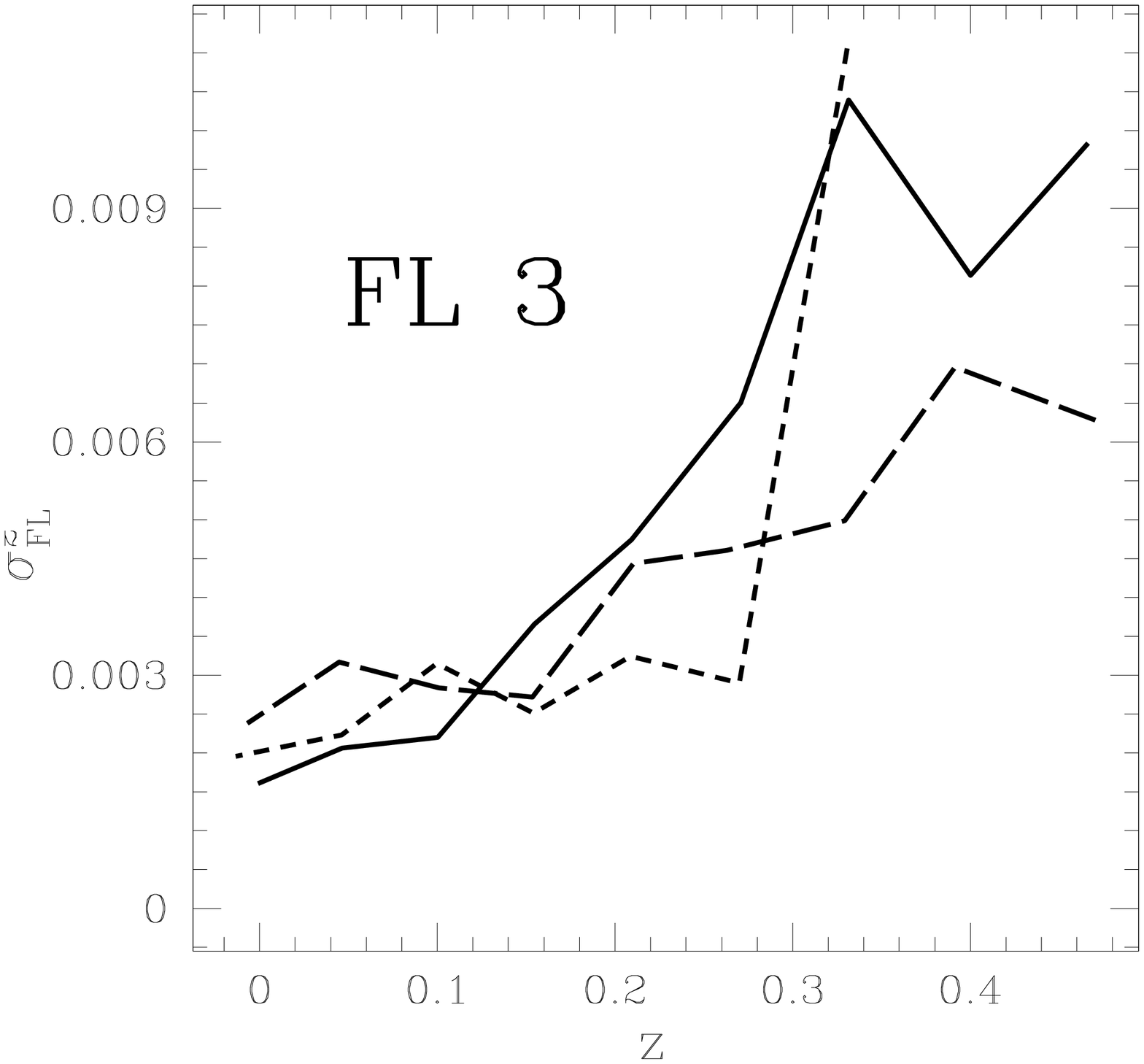}
\caption{Now the fundamental structures are modelled using a line. The mean
scatters around the best-fitting lines are shown for each model as
function of redshift. Linestyles as in Fig.~\ref{fig:fp_both_scatter}.
\label{fig:fl_scatter}}
\end{figure*}
\subsection{Fundamental structures and cluster morphology}
So far our results indicate  that the galaxy clusters are attracted by
a  \emph{quasi-equilibrium state}  mirrored by  fundamental structures
which seem to  be more or less universal  for all cosmological models.
The  evolutions  of the  mean  scatters  show  that the  clusters  are
approaching this quasi-equilibrium state in time in a sort of
\emph{relaxation process}.  In  order to  quantify how  far individual
clusters are away  from this equilibrium state, one  can estimate their
distances  from the  fundamental  plane within  each  of the  parameter
spaces.  The concept  of a  distance within  the parameter  space thus
allows us to measure the inner dynamical state of a cluster.
\\
The physical nature of  this quasi-equilibrium state can be confirmed
if one can show that  different global  cluster  characteristics
are  accompanying this  evolution. An  excellent candidate  is cluster
substructure; indeed, the cosmology-morphology connection is based on
the assumption that  the age of a cluster  and therefore its dynamical
state is reflected by cluster substructure. We have already seen that
 on  average, both the cluster substructure  and the distances
from the fundamental planes are decreasing in time.
\begin{table*} 
\centering    
\caption{Correlations   (Kendall's   $\tau$)   between  the   averaged
substructure  (as measured by  the structure  functions C,  A$_1$, and
S$_1$   for   $r_{\rm   w}=   1.4$   and   a   smoothing   length   of
$\lambda=0.05\mpch$)  and  the   mean  quadratic  scatter  around  the
fundamental lines.   We consider all models and  all parameter spaces.
Similar  results  can  be  obtained  using  the  plane-fitting,
whenever the  scatter around the  plane is decreasing with  time.  The  fact,  that almost  no significant  correlations
occur for the CHDM model, may be explained in part by noticing that we
use fewer  pairs of data points  for the correlation  analysis in this
case.\label{tab:ave_corr}}
\begin{tabular}{llllllll}
model  & FL  & \multicolumn{2}{c}{C} & \multicolumn{2}{c}{A$_1$} & \multicolumn{2}{c}{S$_1$}  \\\hline
 $\lambda =0.05$ &   & $\tau$ &  $p$ & $\tau$ &  $p$ & $\tau$ &  $p$    \\\hline
 CDM & 1  & 0.72 & 0.007   & 0.61 & 0.022   & 0.78 & 0.004  \\\hline
 CDM & 2  & 0.78 & 0.004   & 0.67 & 0.012   & 0.83 & 0.002  \\\hline
 CDM & 3  & 0.78 & 0.004   & 0.67 & 0.012   & 0.83 & 0.002  \\\hline
 CHDM & 1  & 0.62 & 0.051   & 0.62 & 0.051   & 0.71 & 0.024  \\\hline
 CHDM & 2  & 0.24 & 0.453   & 0.24 & 0.453   & 0.33 & 0.293  \\\hline
 CHDM & 3  & 0.24 & 0.453   & 0.24 & 0.453   & 0.33 & 0.293  \\\hline
 \ldm & 1  & 0.72 & 0.007   & 0.83 & 0.002   & 0.83 & 0.002  \\\hline
 \ldm & 2  & 0.50 & 0.061   & 0.61 & 0.022   & 0.61 & 0.022  \\\hline
 \ldm & 3  & 0.61 & 0.022   & 0.72 & 0.007   & 0.72 & 0.007  \\
\end{tabular} 
\end{table*} 
\\
Therefore, we ask  in a first step whether  the \emph{sample-averaged}
substructure  and   the  \emph{sample-averaged   scatter  around  the
fundamental structures} are correlated during their time-evolution. A
basic test relies on  Kendall's $\tau$, a non-parametrical correlation
coefficient~{}\citep{kendall:tau,fritsch:fp}.  In  general, the amount
of  $\tau$  reflects the  strength  of  the  correlations between  two
quantities within a given data set, while the sign of $\tau$ specifies
whether positive or negative  correlations hold among the data points.
Only  values  of  $\tau$  where  $p(\tau)$  (the  probability  of  the
nullhypothesis that  no correlations among  the data points  exist) is
smaller  than $0.05$ evince  a statistically  significant correlation.
The results  shown in Table~\ref{tab:ave_corr} indicate  that, for the
case of the  line-fitting, strong correlations exist for  the CDM and
for the  \ldm model.  For  the CHDM model  we have fewer  redshifts, a
fact, that  in part  may explain  the less meaningful  results.
\begin{figure*}
\centering
\includegraphics[width=5.9cm]{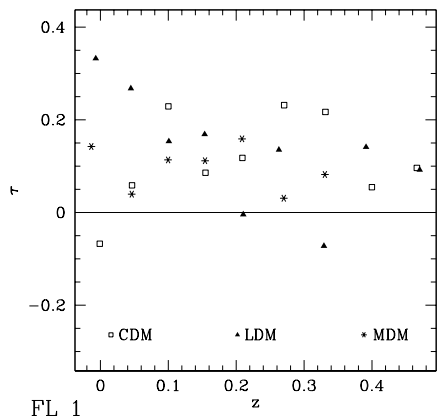}
\caption{We apply a Kendall's test  to relate the substructure and the
dynamics of  individual clusters.  The substructure  is measured using
the clumpiness  (estimated after having employed  a smoothing length  of $0.05\mpch$),
the dynamical  state of  the cluster is  quantified with  the distance
from the first fundamental line. We
show the  results, the Kendall correlation coefficient,  as a function
of redshift  for all  models. Note, that  for significant  results, we
need  a $\tau$  larger than  about  $0.2$ (slightly  depending on  the
number of clusters which is different for the different models.)
\label{fig:l}}
\end{figure*}
\\
To become more specific, we ask  in a second step whether the distance
from the  fundamental plane and the cluster  morphology, quantified by
the structure functions, are  connected for \emph{individual clusters}.
Tentatively  we carried  out  Kendall tests  for  each cluster  sample
relating  the  substructure  parameters  and the  distances  from  the
fundamental  structures.    Since  we  considered   several  structure
functions at different values of the smoothing scale, there is quite a
lot  of freedom.   However, the  results,  as shown,  for example,  in
Fig.~\ref{fig:l}, do not establish a significant correlation between
the  substructure and  the  distance from  the  fundamental plane  for
individual  clusters; there is  no connection  persisting in  time and
throughout all  of the  cosmological models. More  specifically, there
are not many significant correlations to  be found at all; and some of
them   even   turn   out   to   be   anticorrelations   meaning   that
substructure-poor clusters are farer  away from the fundamental plane
than  the substructure-rich  ones. But  there  are also  a number  of
positive  correlations  between  substructure  and  distance  from  an
equilibrium to be discovered for other models at certain redshifts.
\\
 The
lack of  a statistical  significant relation between  substructure and
dynamics  for  individual  clusters   may  have  several  reasons;  in
particular,  a  physical connection  may  be  obscured on  statistical
grounds.   For  example,  if  the  fit of  the  fundamental  plane  is
determined  by a  few clusters  far from  equilibrium, then  the fitted
fundamental structure  is distorted with  respect to the real  one and
the  distances from the  fundamental plane  become distorted  as well.
This can be clarified in future analyses with larger cluster samples.
\\
 It is, however, worth noticing that anticorrelations frequently occur
in cases  where the  mean scatter around  our fits of  the fundamental
structure fails  to decrease during the dynamical  evolution. In other
words, careful  and appropriate fittings which result  in a decreasing
scatter with time reduce in  part the indefinitive results in favour of
a positive  connection between substructure and the  distance from the
fundamental plane.\\ The results on the fundamental structures can be
summarised as follows:
\begin{enumerate}
\item  There is  clear evidence  for  the existence  of a  fundamental
band-like  structure in  all  sorts of  parameter spaces  investigated
here, even for  moderate redshifts ($\lesssim 0.4$). This  band can be
fitted either using a plane or a line.
\item Fitting the fundamental structures appropriately, one can
observe that in most cases the scatter around these structures decreases
with time on the whole.
\item Examining the evolution of the scatter around the fundamental
structures and their morphologies one can specify, that in the first and third parameter space a
line can be seen, whereas the data within the second space
are better fitted by a plane. 
\item  There are  no  large  differences at  redshift  zero among  the
fundamental planes observed in different cosmological models.
\item The  scatters around  the fundamental structures  are comparable
for all cosmological background models investigated here\footnote{This
can     be    seen     from     Figs.~\ref{fig:fp_both_scatter}    and
\ref{fig:fl_scatter}.  A  quantitative analysis  based on the  KS test
shows that indeed the cosmological models are in most cases compatible
with  each   other  regarding  the  distances   from  the  fundamental
lines/planes  for  low  redshifts.    Thus,  it  is  not  possible  to
discriminate between  the different background models  merely by means
of the scatter around the fundamental structures.}.
\item  There  are some  positive  correlations  between the  structure
functions and  the distances from the fundamental  structures (if
appropriately fitted) for the averaged cluster
evolution;  for  individual clusters,  however,  the  results are  not
conclusive.
\end{enumerate}

\section{Conclusions}
\label{sec:con}
The  shape  of  a  body   frequently  is  its  first  property  to  be
recognised. However, when the  body is investigated more closely, more
than  a qualitative  description of  its morphology  is  required. One
wants  to   know,  to  what  extent  its   appearance  reflects  inner
properties,  and  one  wants to  compare  the  shape  of the  body  to
predictions of  analytical models. For both purposes  one needs robust
descriptors of the body's morphology.
This is also  true for galaxy clusters. In  this paper we investigated
both the morphology  and the inner dynamical state  of galaxy clusters
using large samples of simulated  galaxy clusters. In order to measure
the cluster substructure we  employed structure functions based on the
Minkowski valuations;  the inner cluster  state was quantified  by the
distance  from  a  fundamental   band-like  structure  observed  in  a
parameter space  of global cluster descriptors.  The  intention of our
paper  was  twofold: we  first  tested a  new  method  to measure  the
substructure of  galaxy clusters.  On the other  hand, we investigated
how  far  the  cluster  inner  dynamical  state  is  mirrored  by  the
morphology  and  how  different  cosmological  background  models  are
distinguished by means of  the cluster substructure.  \\ Regarding the
morphometry of galaxy clusters,  i.e.  the quantitative description of
their   size,  shape,  connectivity,   and  symmetry,   the  Minkowski
functionals  together  with  the   Querma{\ss}  vectors  allow  for  a
discriminative  and complete  characterisation.  They  are based  on a
number of covariance properties and  thus rest on a solid mathematical
basis.   The  structure functions  constructed  from  the MVs  feature
different  aspects of substructure  successfully.  \\  Employing these
methods   we  showed   that   the  substructure   of  X-ray   clusters
distinguishes  between cosmological  models in  an effective  way.  As
expected theoretically,  the substructure on the whole  is minimal for
the $\Lambda$CDM model and higher for the high-$\om$ models considered
here.   The  power  spectrum  does  not  seem  to  have  a  systematic
influence.  We mainly focused on  simulated X-ray images; but also the
DM substructure  can distinguish between the  cosmological models.  \\
Another  important issue  is the  connection between  substructure and
fundamental   plane   relations.   This   connection   has  not   been
investigated so  far using numerical N-body  simulations.  In general,
the evolution of fundamental plane relations within N-body simulations
has not yet been scrutinised extensively. We could show that there are
stable  fundamental  band-like  structures  within  most  cosmological
models.   Moreover,  we  found  a  positive  correlation  between  the
averaged distance from this  structure (if it is fitted appropriately)
and the sample-averaged structure functions during time for two of our
cosmological models.   For individual clusters, however,  we failed to
produce  definitive  results.   Further  investigations  using  larger
simulations are  in order to  tackle this point.   However, altogether
there are  weak indications that both our  structure functions feature
those aspects of substructure that reflect the inner cluster state and
that  the  fundamental  structures  are  the  imprint  of  a  physical
equilibrium.
\\
These results raise a couple of  new questions: how can we explain the
fundamental   structures?  What   is  the   physical  origin   of  the
degenerated fundamental line?  Are the fundamental bands dependent on
the  environment as  suggested by~\citet{miller:fp}? What is the
precise time evolution of fundamental structures?
\\ 
A  number  of tasks  still  remain  to be  done:  in  this paper,  the
FP-parameters were defined using the three-dimensional clusters. How
significant are  all the  results found here,  when one moves  to more
observation-like  defined quantities?   In our  results one  thing is
 paradoxical:   on  the   one  hand,  the   mean  substructure
discriminates well between  the models, whereas on the  other hand the
mean  scatters around  the fundamental  bands  are  comparable for all sort  of models.    We  conclude that  the morphology is really necessary
to establish  a  connection between  clusters  and the  global
cosmological parameters.

\begin{acknowledgement}
This  work  was  supported  by  the  ``Sonderforschungsbereich  375-95
f{\"u}r Astro-Teilchen\-physik''  der Deutschen Forschungsgemeinschaft
and the Tomalla  foundation, Switzerland.  T.B.  acknowledges generous
support and  hospitality by  the National Astronomical  Observatory in
Tokyo, as  well as hospitality  at Tohoku University in  Sendai, Japan
and Universit\'e de Gen\`eve, Switzerland.  Parts of the codes used to
calculate   the  MVs   are  based   on  the   ``beyond''   package  by
J.~Schmalzing.  Furthermore, we thank  M.~Kerscher for comments on the
manuscript, and  H.~Wagner for useful discussions.   Finally, we thank
the anonymous referee for useful and detailed criticsm.
\end{acknowledgement}

\bibliographystyle{apj}
\bibliography{h2859l}

\end{document}